\documentclass[aps,prx,twocolumn,secnumarabic,amsmath,amssymb]{revtex4-1}

\usepackage{titlesec}
\usepackage{bm}
\usepackage{comment}
\usepackage{verbatim}

\usepackage{graphicx}
\usepackage{subfigure}
\usepackage{tabularx}

\usepackage{color}
\usepackage[colorlinks,bookmarks=false,citecolor=blue,linkcolor=red,urlcolor=blue]{hyperref}

\definecolor{darkred}{rgb}{0.7,0.0,0.0}

\definecolor{darkblue}{rgb}{0,0.02,0.45}

\definecolor{darkgreen}{rgb}{0.02,0.45,0.0}

\definecolor{violet}{rgb}{0.8,0.2,0.6}

\def\be{\begin{equation}}
\def\ee{\end{equation}}
\def\bea{\begin{eqnarray}}
\def\eea{\end{eqnarray}}

\def\vec{\mathbf}
\def\bs{\boldsymbol}
\def\mc{\mathcal}

\usepackage{makecell}
\usepackage{boldline}
\setcellgapes{3pt}

\usepackage{times}

\begin{document}

\date{\today}

\title{Quantum Spin Liquid in the semiclassical regime}

\author{Ioannis Rousochatzakis}
\affiliation{School of Physics and Astronomy, University of Minnesota, Minneapolis, MN 55455, USA}

\author{Yuriy Sizyuk}
\affiliation{School of Physics and Astronomy, University of Minnesota, Minneapolis, MN 55455, USA}

\author{Natalia B. Perkins}
\affiliation{School of Physics and Astronomy, University of Minnesota, Minneapolis, MN 55455, USA}

\begin{abstract}
Quantum spin liquids have been at the forefront of correlated electron research ever since their original proposal in 1973, and the realization that they belong to the broader class of intrinsic topological orders, along with the fractional quantum Hall states. According to received wisdom, quantum spin liquids can arise in frustrated magnets with low spin $S$, where strong quantum fluctuations act to destabilize conventional, magnetically ordered states. Here we present a magnet that has a $Z_2$ quantum spin liquid ground state already in the semiclassical, large-$S$ limit. The state has both topological and symmetry related ground state degeneracy, and two types of gaps, a `magnetic flux' gap that scales linearly with $S$ and an `electric charge' gap that drops exponentially in $S$. The magnet is described by the spin-$S$ version of the spin-1/2 Kitaev honeycomb model,  which has been the subject of intense studies in correlated electron systems with strong spin-orbit coupling, and in optical lattice realizations with ultracold atoms. The results apply to both integer and half-integer spins. 
\end{abstract}

\maketitle

\pagebreak

Quantum spin liquids (QSLs) describe systems that evade magnetic long-range order down to zero temperature, and manifest a number of remarkable phenomena, 
such as topological ground state degeneracies, emergent gauge fields, and fractional excitations with non-trivial statistics.~\cite{Anderson1973,FazekasAnderson74,Kalmeyer1987,Wen1989,Moessner2001, Kitaev2003,Kitaev2006,Levin2005,Balents2010,Savary2016,Zhou2017} 
The rich phenomenology of QSLs derives from an intrinsic tendency to form massive quantum superpositions of local, `product-like' wavefunctions. Notable examples are the resonating valence bond (RVB) state,~\cite{Anderson1973,RokhsarKivelson,Moessner2001,Misguich02,Hao14} the gapped QSL of the Toric code,~\cite{Kitaev2003} and the gapless QSL phase of the spin-1/2 Kitaev honeycomb model.~\cite{Kitaev2006}

Typically, such massive superpositions arise in frustrated magnets with low spin $S$, which ideally have an infinite number of competing states and a strong tunneling between them.~\cite{HFMBook} 
Here we show that the spin-$S$ version of the celebrated Kitaev honeycomb model is a topological $Z_2$ QSL  already in the semiclassical  limit. Specifically, the leading semiclassical fluctuations give rise to an effective low-energy description in terms of a pseudospin-1/2 Toric code.~\cite{Kitaev2003} 
The `magnetic flux' term of the Toric code arises from the zero-point energy of spin waves above the classical ground states, while the `electric charge' term stems from the tunneling between different classical states. 
The ensuing $Z_2$ QSL lives on top of a honeycomb superlattice of `frozen' spin dimers,~\cite{Baskaran08} which take only two possible configurations, instead of $(2S\!+\!1)^2$. These two states are the pseudospin-1/2 degrees of freedom of the Toric code.
The frozen dimer pattern breaks translational symmetry, so the QSL possesses an extra degeneracy associated to symmetry breaking, besides the topological one.

The $Z_2$ gauge structure is not an emergent property of the low-energy sector of the problem, but descends from the gauge structure of the original spin-$S$ model, which was discovered in a seminal study by Baskaran, Sen and Shankar (BSS).~\cite{Baskaran08} 
As such, the gauge structure is not only present in the low-energy sector, but also in the single-particle, spin-wave excitation channel, which we analyze in detail beyond the quadratic level. 

The large-$S$ description breaks down around $S\!\sim\!3/2$. For lower $S$, tunneling processes that shift the dimer positions become quickly relevant and compete with the `freezing' energy scale $\delta E_\text{f}$. Including these processes leads to a picture of `decorated quantum dimers', where both the dimer positions and the orientations of the two spins in each dimer are allowed to resonate. 
The ensuing picture for $S\!=\!1$ in terms of another type of spin liquid will be discussed.

\begin{figure*}[!t] 
\vspace{-0.25cm}
\includegraphics[width=0.75\textwidth,angle=0,clip=true,trim=0 0 0 0]{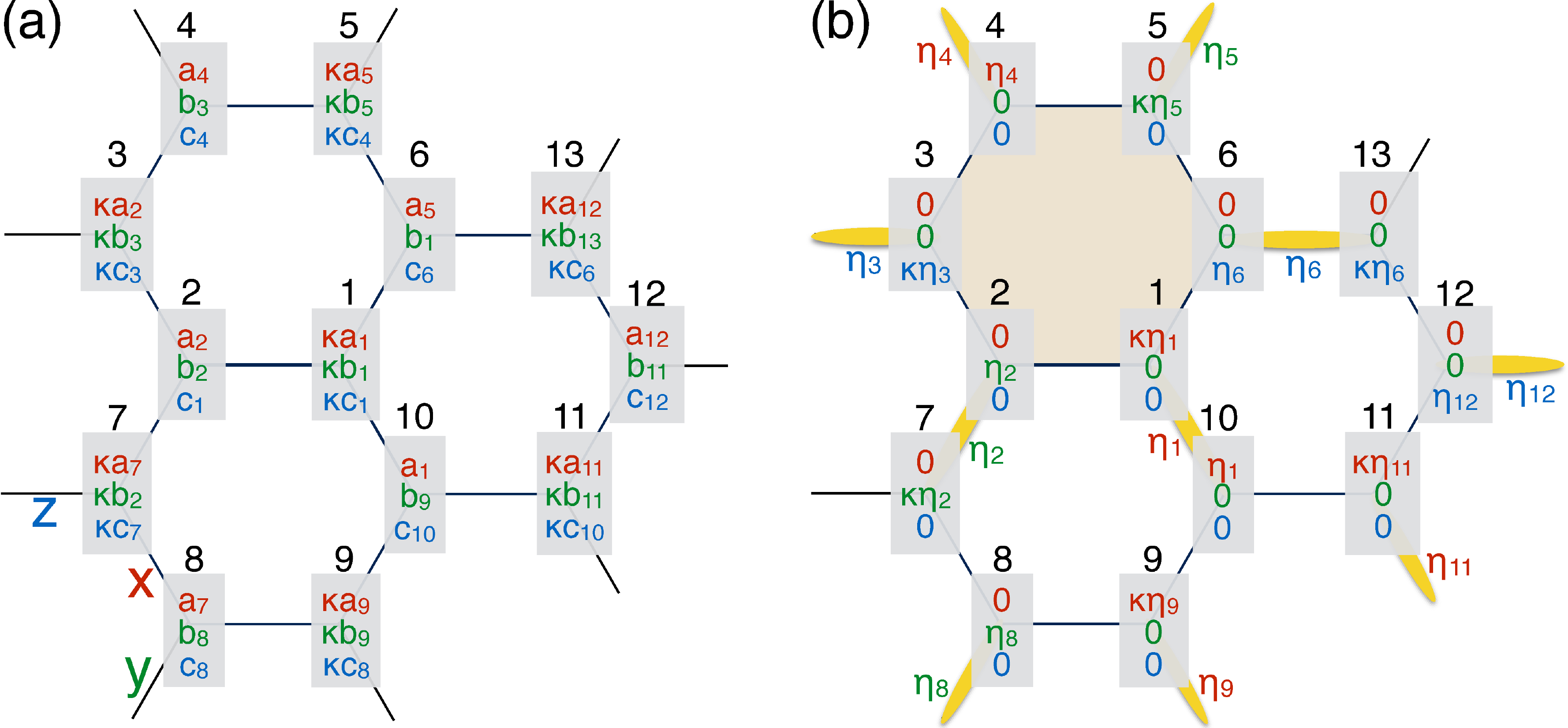}
\vspace{-0.25cm}
\caption{(a) Classical ground states of the Kitaev model. Here $\kappa\!=\!-\text{sgn}(K)$, $\vec{S}_i\!=\!(a_i,b_i,c_i)$ or $\kappa (a_i,b_i,c_i)$ if $i$ belongs to the A or B sublattice, and $a_i^2\!+\!b_i^2\!+\!c_i^2\!=\!S^2$. 
(b) The Cartesian states of BSS map to dimer coverings, with (yellow) dimers representing satisfied bonds. The spin orientation of each dimer is described by an Ising-like variable $\eta\!=\!\pm1$. The shaded hexagon has the shortest loop with no dimers.}\label{fig:ClassGSs}
\end{figure*}

\vspace*{0.1cm}
\noindent {\bf Results}\\
\noindent{\bf Model \& classical ground states.} 
The spin-$S$ Kitaev model on the honeycomb lattice is described by the Hamiltonian \
\be\label{eq:Kit}
\mc{H} = K \Big( \!\!\sum_{\langle ij\rangle \in \text{`x'}}\!\!\!\! S_i^xS_j^x
+\!\!\!\!\sum_{\langle ij\rangle \in \text{`y'}} \!\!\!\!S_i^yS_j^y
+\!\!\!\!\sum_{\langle ij\rangle \in \text{`z'}} \!\!\!\!S_i^zS_j^z\Big)\,,
\ee
where `x', `y' and `z' denote the three orientations of nearest neighbor (NN) bonds, see Fig.~\ref{fig:ClassGSs}~(a), and $K$ is the coupling constant. Note that there is a four-sublattice duality transformation~\cite{ioannisK1K2} that maps the positive $K$ to the negative $K$ model, but we shall discuss the general case here for completeness. We shall also define $\kappa\!=\!-\text{sgn}(K)$. 

The classical ground states of this model were first analyzed by BSS.~\cite{Baskaran08} There, the authors identified an infinite number of so-called `Cartesian' states, which map to dimer coverings of the honeycomb lattice, modulo a factor of two for the orientation of the two spins per dimer. They further showed that the Cartesian states are connected to each other by continuous valleys of other ground states, leading to a huge ground state degeneracy. Soon after, Chandra, Ramola and Dhar~\cite{Chandra2010} showed that the manifold actually consists of infinitely more ground states and possesses an emergent gauge structure that leads to power-law correlations.
 
The crucial aspect of the present study is the use of a convenient parametrization of the classical ground state manifold, which reveals the topological terms arising from quantum fluctuations in an explicit way.  This parametrization is shown in Fig.~\ref{fig:ClassGSs}~(a). We denote the two sublattices of the honeycomb by A and B.  Next, we parametrize each spin as $\vec{S}_i\!=\!(a_i,b_i,c_i)$ or $\kappa(a_i,b_i,c_i)$ for $i\!\in\!\text{A}$ or B, respectively, and $a_i^2\!+\!b_i^2\!+\!c_i^2\!=\!S^2$. Then, for every pair of NN sites, ${\bf S}_i$ and ${\bf S}_j$, we can minimize their mutual interaction by requiring that $a_i\!=\!a_j$ or $b_i\!=\!b_j$ or $c_i\!=\!c_j$, if the two sites share, respectively, an `x' or `y' or `z' type of bond. To see if the ensuing states are ground states we check that they saturate the lower bound of the energy per site, $E_{\text{min}}/N\!=\!-|K| S^2/2$.~\cite{Baskaran08} Indeed, the energy from the three bonds emanating from any site $i$ add up to $-|K| (a_i^2\!+\!b_i^2\!+\!c_i^2)\!=\!-|K|S^2$. And since each bond is shared by two sites, these configurations saturate the lower bound and are therefore ground states. The Cartesian states of BSS arise by keeping only one component of $(a_i,b_i,c_i)$ finite, and equal to $\eta_i S$, where $\eta_i\!=\!\pm 1$. Modulo these Ising-like variables, the Cartesian states map to dimer coverings of the lattice [Fig.~\ref{fig:ClassGSs}~(b)]. There are $(1.381)^{N/2}$ coverings,~\cite{Wu2006,Baxter1970,Kasteleyn1963} and $(1.662)^N$ Cartesian states in total.~\cite{Baskaran08} 

The semiclassical analysis leading to the Toric code proceeds in three steps. The first is to show that fluctuations select the Cartesian over the non-Cartesian states, which identifies the positions and spin orientations of the dimers as the relevant degrees of freedom.
In the second step, which was carried out by BSS,~\cite{Baskaran08} fluctuations freeze the positions of the dimers to a given pattern, leaving their spin orientation as the only relevant degrees of freedom below the associated  freezing energy scale $\delta E_{\text{f}}$. At this point, our parametrization reveals, in addition, a topological structure that was not noticed previously.
The third step is to include quantum-mechanical tunneling between states with different orientations of the dimers. 

\vspace*{0.1cm}\noindent
{\bf Order-by-disorder I: Selection of Cartesian states.} 
The first crucial ingredient of the effective description in terms of dimers is to show that fluctuations select the Cartesian over the non-Cartesian states. BSS made this hypothesis based on an analogy to a related 1D problem. Here we prove it by real space perturbation theory (RSPT).~\cite{Lindgard1988,Long1989,Heinila1993,Mike2014} We introduce local frames $({\bf u}_i,{\bf v}_i,{\bf w}_i)$, with $\vec{w}_i$ along the classical directions, and write ${\bf S}_i \!=\! S_i^{w} {\bf w}_i\!+\!S_i^{u} {\bf u}_i\!+\!S_i^{v} {\bf v}_i$. Then we split $\mc{H}$ into a diagonal part $\mc{H}_0\!=\!h\sum_i (S\!-\!S_i^w)$, describing fluctuations in the local field $h\!=\!K S$, and a perturbation $\mc{V}\!=\!\mc{H}\!-\!\mc{H}_0$, which couples fluctuations on different sites.
The essential physics is captured by the leading, short-wavelength corrections from second-order RSPT. The three types of bonds, say $(1\text{-}10)$, $(1\text{-}6)$ and $(1\text{-}2)$  of Fig.~(\ref{fig:ClassGSs}), give $\delta E_{1,10}\!=\!\xi(1\!-\!\widetilde{a}_1^2)^2$, $\delta E_{1,6}\!=\!\xi(1\!-\!\widetilde{b}_1^2)^2$, $\delta E_{1,2}\!=\!\xi(1\!-\!\widetilde{c}_1^2)^2$,where $\xi\!=\!-|K| S/8$ and $(\widetilde{a}_i,\widetilde{b}_i,\widetilde{c}_i)\!=\!(a_i,b_i,c_i)/S$. Using the spin length constraints and disregarding overall constants, gives the anisotropy term
\be
\delta E_{\text{ani}} = -(|K| S/16) \sum\nolimits_i (\widetilde{a}_i^4+\widetilde{b}_i^4+\widetilde{c}_i^4)\,,
\ee
similar to the one found in \cite{IoannisGamma,Avella2015}. This anisotropy selects the Cartesian states, confirming the hypothesis of BSS.~\cite{Baskaran08}

\begin{figure*}[!t] 
\vspace*{-0.25cm}
\includegraphics[width=0.75\textwidth,angle=0,clip=true,trim=0 0 0 0]{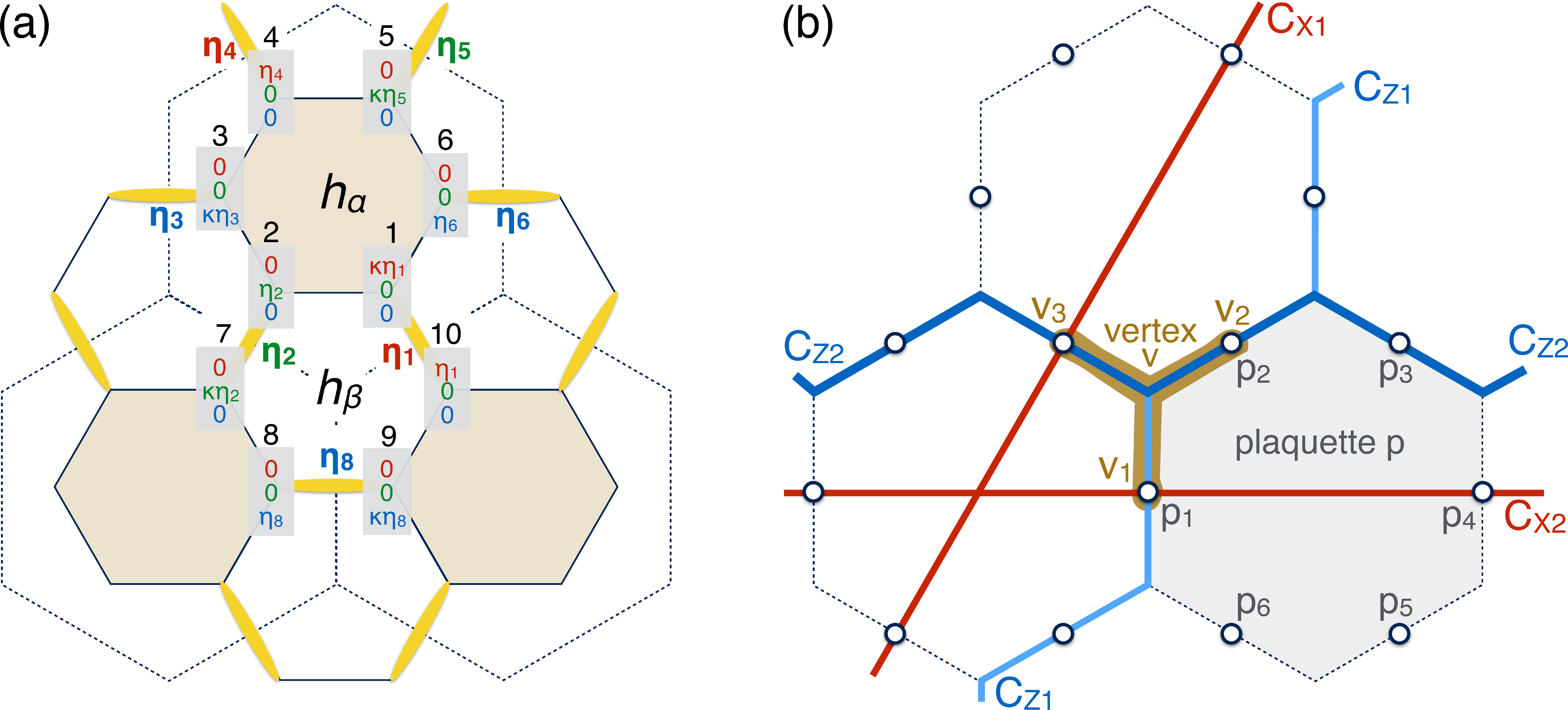}
\vspace*{-0.25cm}
\caption{(a) `Star' dimer (yellow) pattern selected from spin waves.~\cite{Baskaran08} The $\eta$'s describe the orientation of the two spins per dimer. They sit at the middle of the bonds of a honeycomb superlattice (dashed).
(b) The resulting Toric code description of Eq.~(\ref{eq:TC1}) on the honeycomb superlattice. The three- and six-body operators ${\bf A}_v$ and ${\bf B}_p$ of Eq.~(\ref{eq:TC1}) are defined on vertices $v$ and plaquettes $p$ of the superlattice. For the torus geometry, C$_{\text{X1}}$ and C$_{\text{X2}}$ (similarly for C$_{\text{Z1}}$ and C$_{\text{Z2}}$) are non-contractible loops that wrap the system in different directions.
}\label{fig:star&Toric}
\end{figure*}

\vspace*{0.1cm}\noindent
{\bf Order-by-disorder II: Dimer freezing.}  
Next, we discuss the lifting of the degeneracy within the infinite sub-manifold of Cartesian states, starting with the corrections from spin waves. As shown by BSS, i) the linear spin wave Hamiltonian splits into non-interacting modes propagating along loops without dimers, and ii) the minimum zero-point energy arises by maximizing the number of the shortest such `empty' loops, like the  shaded hexagon of Fig.~\ref{fig:ClassGSs}~(b). This gives the `star' or `columnar' dimer pattern of Fig.~\ref{fig:star&Toric}\,(a), which is known from the context of the quantum dimer model and the frustrated Heisenberg model on the honeycomb lattice.~\cite{Moessner2001,Albuquerque2011,Schlittler2015b} 
In this pattern, the only dynamical degrees of freedom remaining are the Ising-like variables $\eta\!=\!\pm1$, which specify the direction of the two spins shared by each given dimer.

The physics of the dimer freezing is actually more involved from what is predicted from the linear spin-wave theory, but let us postpone this discussion for later and focus on the spin states associated to the `star' pattern.
There are three ways to place this pattern in the lattice and each dimer has two configurations, so at first sight, the number of selected spin states is $3\!\times\!2^{N/2}$.
BSS showed, however, that the minimum zero-point energy is associated with spin-wave modes that have antiperiodic boundary conditions (ABC) around the empty hexagons, which reduces the number of states to $3\!\times\!2^{N/3}$.

However, this is not the full story yet. It turns out that the boundary condition on the spin wave modes actually endows the selected manifold with a topological magnetic flux term (and, in particular, the above number of states has to be multiplied $2^{2g-1}$, where $g$ is the genus of the system). To see this, we repeat the spin wave analysis using our $\eta$-parametrization. We begin by rewriting $\mc{H}$ in the local frame. Let us take the empty hexagon $h_\alpha$ of Fig.~\ref{fig:star&Toric}\,(a) and choose ${\bf u}_i$ and ${\bf v}_i$ in the following way (and similarly for every other empty hexagon):
\be\label{eq:localaxes1}
\begin{array}{lll}
\vec{w}_1\!=\!\kappa \eta_1 {\bf x}, 	&\vec{u}_1\!=\!-\kappa \eta_1 {\bf z},	&\vec{v}_1\!=\!{\bf y},\\
\vec{w}_2\!=\! \eta_2 {\bf y}, 		&\vec{u}_2\!=\! -\kappa\eta_1 {\bf z},		&\vec{v}_2\!=\!-\kappa \eta_1\eta_2 {\bf x},\\
\vec{w}_3\!=\!\kappa \eta_3 {\bf z}, 	&\vec{u}_3\!=\!\eta_1\eta_2\eta_3 {\bf y}, 	&\vec{v}_3\!=\!-\kappa \eta_1\eta_2 {\bf x},\\
\vec{w}_4\!=\!\eta_4 {\bf x}, 		&\vec{u}_4\!=\!\eta_1\eta_2\eta_3 {\bf y}, 		&\vec{v}_4\!=\!\eta_1\eta_2\eta_3\eta_4{\bf z},\\
\vec{w}_5\!=\!\kappa \eta_5 {\bf y}, 	&\vec{u}_5\!=\!-\kappa \eta_1\eta_2\eta_3\eta_4\eta_5 {\bf x}, 	&\vec{v}_5\!=\!\eta_1\eta_2\eta_3\eta_4{\bf z},\\
\vec{w}_6\!=\! \eta_6 {\bf z}, 		&\vec{u}_6\!=\! -\kappa \eta_1\eta_2\eta_3\eta_4\eta_5 {\bf x},		&\vec{v}_6\!=\!-\kappa B_{h_\alpha} {\bf y},
\end{array}
\ee
where the product of the six $\eta$-variables on empty hexagons,   
\be
B_{h_\alpha}\!=\!\eta_1\eta_2\eta_3\eta_4\eta_5\eta_6,
\ee
is the magnetic flux that plays a central role in the following. 
With the above choice of the local frames, the couplings between empty hexagons map to terms of the type $\kappa S_i^w S_j^w$. For example, $S_1^x S_{10}^x\!\mapsto\!\kappa S_1^w S_{10}^w$.  
On the other hand,  the intra-hexagon terms map as follows 
\bea
\begin{array}{ll}
S_1^z S_2^z \mapsto S_1^u S_2^u,~ &
S_2^x S_3^x \mapsto S_2^v S_3^v,\\
S_3^y S_4^y \mapsto S_3^u S_4^u,~ &
S_4^z S_5^z \mapsto S_4^v S_5^v,\\
S_5^x S_6^x \mapsto S_5^u S_6^u,~ & 
S_6^y S_1^y \mapsto -\kappa B_{h_\alpha} S_6^v S_1^v\,.
\end{array}
\eea
Thus, in the rotated frame, the only dependence of the Hamiltonian on $\eta$'s is via the products $\{B_{h_\alpha}\}$ on the empty hexagons $\{h_\alpha\}$. And since the choice of the local frame does not alter the physics, it follows that  classical states that belong to the `star' pattern  of Fig.~\ref{fig:star&Toric}\,(a) and have the same $\{B_{h_\alpha}\}$ share the same semiclassical spin wave spectrum, at all orders in $1/S$. (We shall see below that this property reflects a local gauge symmetry of the model.~\cite{Baskaran08}.) The same is true for the renormalization of the ground state energy and therefore the order-by-disorder effect. 
Let us show the latter explicitly and we shall return to the spin-wave modes further below. 

\begin{figure}[!b] 
\vspace*{-0.25cm}
\includegraphics[width=0.4\textwidth,angle=0,clip=true,trim=0 0 0 0]{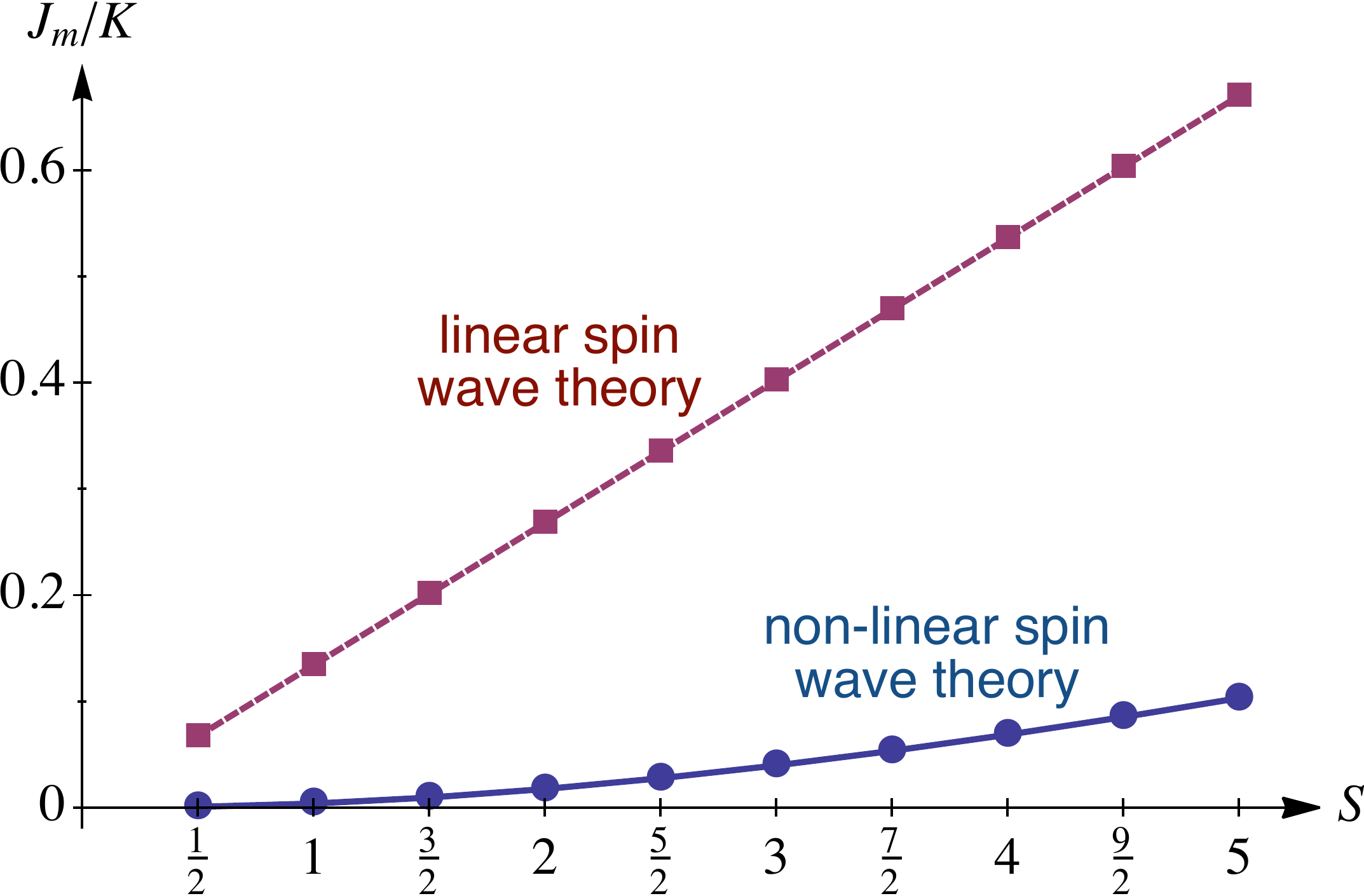}
\vspace*{-0.25cm}
\caption{The coupling $J_m$ as a function of $S$, extracted from linear (dashed) and non-linear (solid) spin wave theory.}\label{fig:JmvsS}
\end{figure}

\begin{figure*}[!t] 
\vspace{-0.25cm}
\includegraphics[width=0.9\textwidth,angle=0,clip=true,trim=0 0 0 0]{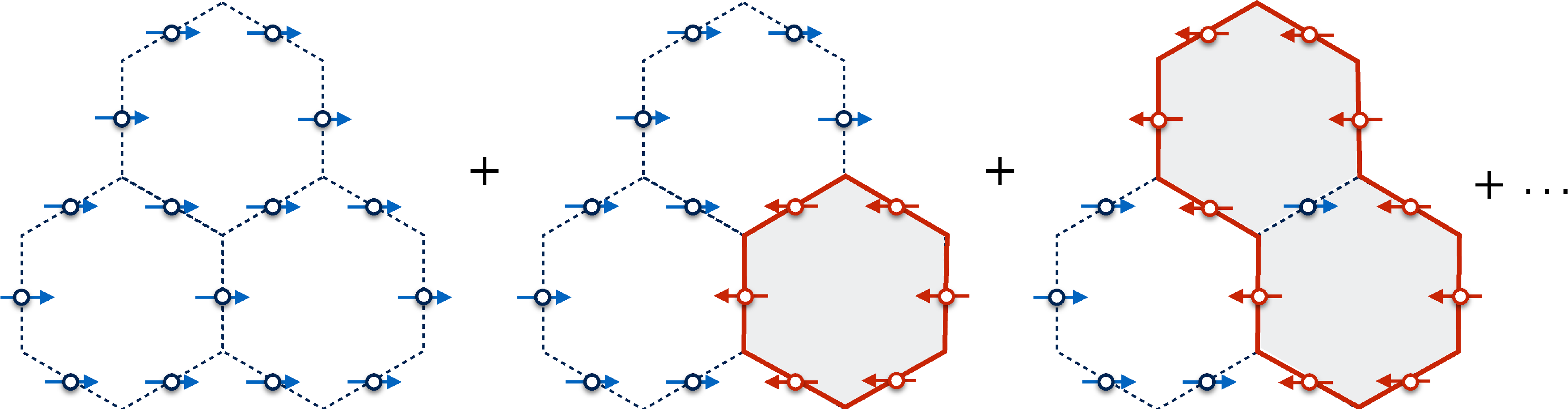}
\vspace*{-0.25cm}
\caption{The quantum spin liquid ground states of the Toric code (\ref{eq:TC1}) correspond to massive, equal-amplitude superpositions of all possible loops of spins (red solid lines) pointing along $-{\bf x}$ (red arrows), on top of a FM background of spins pointing along $+{\bf x}$ (blue arrows).}\label{fig:GS}
\end{figure*}

We introduce the usual Holstein-Primakoff bosons $a_i$ via the transformation,~\cite{HP1940} $S_i^\dagger\!=\!S_i^u+i S_i^v\!=\!(2S\!-\!a_i^\dagger a_i)^{1/2} a_i$ and $S_i^w\!=\!S\!-\!a_i^\dagger a_i$. 
To order $\mc{O}(S)$, empty hexagons decouple, leading to a quadratic, six-site boson problem, with two sublattices and periodic (PBC) or antiperiodic (ABC) boundary conditions, for $\kappa B_{h_\alpha}\!=\!-1$ or $1$, respectively. So the BSS result that ABC give the lowest zero-point energy amounts to imposing $\kappa B_{h_\alpha}\!=\!1$ for all empty hexagons $h_\alpha$. 
More explicitly, by combining the energies~\cite{Baskaran08} $\delta E_{\text{PBC}}$ and $\delta E_{\text{ABC}}$ for PBC and ABC, respectively, we get the contribution to the zero-point energy from $h_\alpha$,
\be\label{eq:pot}
\delta E(h_\alpha) = c + J_m \eta_1\eta_2\eta_3\eta_4\eta_5\eta_6 = c + J_m B_{h_\alpha}\,,
\ee
where $c\!=\!\frac{\delta E_{\text{PBC}}\!+\!\delta E_{\text{ABC}}}{2}$ and $J_m\!=\!\frac{\delta E_{\text{PBC}}\!-\!\delta E_{\text{ABC}}}{2}$. This shows that the corrections to the ground state energy depend explicitly on the fluxes $\{B_{h_\alpha}\}$, and that states with the same set of fluxes have the same zero-point energy.

The linear spin-wave theory of BSS~\cite{Baskaran08} gives $\delta E_{\text{PBC}}\!=\!2|K|S$ and $\delta E_{\text{ABC}}\!=\!\sqrt{3}|K|S$, and so $J_m\!=\!\frac{2\!-\!\sqrt{3}}{2}K S$. However, as shown in Fig.~\ref{fig:JmvsS} and emphasized below, the linear theory overestimates $|J_m|$ strongly due to the presence of a large percentage (four out of six) of `spurious' zero modes. 

We are now ready to identify the first crucial ingredient of the Toric code description announced above. The $\eta_i$ variables live on the midpoints of the bonds of a honeycomb superlattice (Fig.~\ref{fig:star&Toric}\,(a)), and Eq.~(\ref{eq:pot}) tells us that promoting these variables to Pauli matrices ${\bs \eta}_i^z$ leads to the magnetic flux term of the Toric code~\cite{Kitaev2003} on this superlattice.

\vspace*{0.1cm}\noindent
{\bf Order-by-disorder III: Tunneling.} 
The second ingredient of the Toric code, the electric charge term, stems from processes that flip the three $\eta$'s around a vertex of the superlattice.  Let us take, e.g., the spin coherent state of the $h_\beta$ hexagon of Fig.~\ref{fig:star&Toric}\,(a),
\be
|h_\beta\rangle\!=\!|\kappa\eta_8{\bf z}\rangle_{9}  |\eta_8{\bf z}\rangle_8  |\kappa\eta_2{\bf y}\rangle_{7} |\eta_2{\bf y}\rangle_2  |\kappa\eta_1{\bf x}\rangle_1 |\eta_1{\bf x}\rangle_{10}\,.
\ee 
The leading processes that transform this state to its time reversed state $|\bar{h}_\beta\rangle$, with $\eta_1$, $\eta_2$ and $\eta_8$ flipped, appear in $(6S)$-th order of RSPT, with $\mc{V}\!=\!K(S_{7}^xS_{8}^x\!+\!S_{9}^yS_{10}^y\!+\!S_{1}^zS_{2}^z)$. The corresponding off-diagonal matrix element $J_e$ of the resulting effective Hamiltonian $\mc{H}_{\text{eff}}$ depends, unlike $J_m$, on the choice of the local axes $({\bf u}_i, {\bf v}_i)$. Here we fix $J_e$ to be a real number by choosing the local axes such that $\mc{V}\!\mapsto\!K(S_{7}^uS_{8}^u\!+\!S_{9}^uS_{10}^u\!+\!S_{1}^uS_{2}^u)$.  Following e.g., the steps of the Supplemental Material of \cite{IoannisGamma}, we get  
\bea\label{eq:Tunn}
\langle \bar{h}_\beta|\mc{H}_{\text{eff}}|h_\beta\rangle\!\equiv\!
J_e\!=\!  3 \times 2^{5-18S} S^{5-6S} [(2S\!-\!1)!]^3 K\,.~~~
\eea
In the language of the ${\bs \eta}$ operators, this matrix element is represented by $J_e {\bs \eta}_1^x{\bs \eta}_2^x{\bs \eta}_8^x$, which involve the three $\eta$'s around the vertex that sits at the center of $h_\beta$ (see Fig.~\ref{fig:star&Toric}\,(a)).

\vspace*{0.1cm}\noindent
{\bf Toric code model.} Collecting the potential energy (disregarding $c$) and the tunneling terms above gives [see Fig.~\ref{fig:star&Toric}\,(b)]: 
\bea\label{eq:TC1}
\mc{H}_{\text{eff}} &=& J_e \sum\nolimits_{v} {\bs \eta}_{v_1}^x {\bs \eta}_{v_2}^x {\bs \eta}_{v_3}^x
+ J_m \!\sum\nolimits_{p}\! {\bs \eta}^z_{p_1} \!\cdots {\bs \eta}^z_{p_6}, \nonumber\\
&\equiv& J_e \sum\nolimits_{v} {\bf A}_v + J_m \sum\nolimits_{p} {\bf B}_p~,
\eea
where $v$ and $p$ label, respectively, the vertices and the plaquettes of the honeycomb superlattice. In terms of the original lattice, the former sit at the centers of non-empty hexagons of type $h_\beta$, while the latter enclose the empty hexagons of type $h_\alpha$. Essentially then, $v$ and $p$ label $h_\beta$ and $h_\alpha$, respectively.

The remarkable properties of the model (\ref{eq:TC1}) stem from the relations  ${\bf A}_v^2\!=\!{\bf B}_p^2\!=\!1$ and the fact that $\{{\bf A}_v, {\bf B}_p, \mc{H}_{\text{eff}}\}$ is a set of mutually commuting operators.~\cite{Kitaev2003}
This model is a $Z_2$ lattice gauge theory,~\cite{Wegner1971,Kogut1979} with the local gauge transformations generated by ${\bf A}_v$. 
In the following, we discuss the most important properties~\cite{Kitaev2003,Levin2005,Savary2016} of the Toric code. Without loss of generality, we will consider the $K\!<\!0$ case, where both $J_m$ and $J_e$ are negative.

\vspace*{0.1cm}\noindent
{\bf Topological sectors.} 
On a torus, $\prod_v {\bf A}_v\!=\!\prod_p {\bf B}_p \!=\!1$ and so there are  $N_v\!=\!2^{N/3-1}$  and $N_p\!=\!2^{N/6-1}$ independent choices of ${\bf A}_v$ and ${\bf B}_p$, respectively, leading to $2^{N/2-2}$ states. So the quantum numbers $\{A_v, B_p\}$ do not exhaust all $2^{N/2}$ states of ${\bs \eta}$'s. The missing quantum numbers are provided by the nonlocal operators ${\bf X}_1\!=\!\prod_{\text{C}_{\text{X1}}} \!\!{\bs \eta}^x$ and ${\bf X}_2\!=\!\prod_{\text{C}_{\text{X2}}}\!\! {\bs \eta}^x$, defined on the non-contractible loops C$_{\text{X1}}$ and C$_{\text{X2}}$ of Fig.~\ref{fig:star&Toric}\,(b). These operators commute with ${\bf A}_v$ and ${\bf B}_p$, and with each other, and in addition ${\bf X}_1^2\!=\!{\bf X}_2^2\!=\!1$. The quantum numbers $\{ A_v, B_p, X_1, X_2\}$ then exhaust the Hilbert space of ${\bs \eta}$'s.

\vspace*{0.1cm}\noindent
{\bf Ground states.} 
The ground states have $A_v\!=\!B_p\!=\!1$, $\forall v, p$. On a torus, there are four such states, which correspond to the choices of the winding numbers $X_1$ and $X_2$. One of them is 
\be\label{eq:11}
|X_1\!=\!1,X_2\!=\!1\rangle\!=\!\mc{N} \!\prod\nolimits_{p}\big(1+{\bf B}_p\big) ~
|\text{FM}_x\rangle~,
\ee 
where $\mc{N}$ is a normalization factor, and $|\text{FM}_x\rangle\!=\!|\!\!\rightarrow\cdots\rightarrow\rangle$ is the fully polarized state along ${\bf x}$, which has $A_v\!=\!1$, $\forall v$. Expanding the product over $(1\!+\!{\bf B}_p)$ shows that this state is the equal amplitude superposition of all possible loops of overturned spins (spins pointing along $-{\bf x}$, which correspond to electric flux lines) on top of the FM background, see Fig.~\ref{fig:GS} and \cite{Kitaev2006,Levin2005}.
The remaining three ground states of the Toric code, $|X_1,X_2\rangle\!=\!|\text{-}1,1\rangle$, $|1,\text{-}1\rangle$ and $|\text{-}1,\text{-}1\rangle$, arise by replacing the reference state $|\text{FM}_x\rangle$ in (\ref{eq:11}) with ${\bf Z}_2|\text{FM}_x\rangle$, ${\bf Z}_1|\text{FM}_x\rangle$ and ${\bf Z}_1{\bf Z}_2|\text{FM}_x\rangle$, respectively, where ${\bf Z}_1\!=\!\prod_{C_{Z_1}} \!\!{\bs \eta}^z$ and ${\bf Z}_2\!=\!\prod_{C_{Z_2}}\!\! {\bs \eta}^z$, defined along $C_{Z_1}$ and $C_{Z_2}$ of Fig.~\ref{fig:star&Toric}\,(b). These operators flip $X_2$ and $X_1$, respectively, because of  the anti-commutation relations $\{{\bf Z}_1,{\bf X}_2\}\!=\!0$ and $\{{\bf Z}_2,{\bf X}_1\}\!=\!0$.

Importantly, the ground state sector of the original Kitaev spin model is 12-fold and not 4-fold degenerate, because there are three ways to place the dimer pattern of Fig.~\ref{fig:star&Toric}\,(a) into the lattice and each sector has its own Toric code description.

\vspace*{0.1cm}\noindent
{\bf Excitations of Toric code (\ref{eq:TC1}).} 
The elementary excitations are pairs of static charges (vertices with $A_v\!=\!-1$), or pairs of static fluxes (plaquettes with $B_p\!=\!-1$). Their energy is $\Delta_e\!=\!4|J_e|$ and $\Delta_m\!=\!4|J_m|$, respectively. So $\Delta_m$ scales roughly linearly with $S$ (see Fig.~\ref{fig:JmvsS}), whereas $\Delta_e$ is exponentially small in $S$, as follows from Eq.~(\ref{eq:Tunn}), and practically vanishes for $S\!\ge\!1$ and realistic values of $K$. These excitations describe deconfined particles (the energies do not depend on the distance between the charges or fluxes) and they also carry nontrivial mutual statistics.~\cite{Kitaev2003}

\vspace*{0.1cm}\noindent
{\bf Origin of gauge structure \& BSS fluxes.} 
The local $Z_2$ gauge symmetry of (\ref{eq:TC1}) is not an emergent property, but descends from the $Z_2$ gauge structure of the original spin-$S$ model, discovered by BSS.~\cite{Baskaran08} This structure stems from the presence of local conserved operators defined on the hexagons of the original lattice, which are called BSS fluxes in the following. For the $h_\beta$ hexagon of Fig.~\ref{fig:star&Toric}\,(a), the BSS flux operator reads:
\be
{\bf W}_{\text{BSS}}(h_\beta)\!=\!\exp[ i \pi (S_{9}^x +S_8^y +S_{7}^z +S_2^x +S_1^y +S_{10}^z)]\,.
\ee 
Now, the BSS fluxes on non-empty hexagons have the same effect as the ${\bf A}_v$ operators, e.g.\! ${\bf W}_{\text{BSS}}(h_\beta) |h_\beta\rangle\!\to\! |\bar{h}_\beta\rangle$ (modulo some prefactor, see (\ref{eq:Wbeta}) below). So the local gauge symmetry of (\ref{eq:TC1}) indeed descends from that of the full model.

Let us now examine the ground state BSS flux pattern.
Unlike the original classical states associated with the `star' pattern, where only the empty hexagons have well defined ${\bf W}_{\text{BSS}}$,~\cite{Baskaran08} the QSL ground states of (\ref{eq:TC1}) have well defined ${\bf W}_{\text{BSS}}$ on all hexagons. Indeed, using the same choice of local axes as the ones used above for the tunneling we find:
\be\label{eq:Wbeta}
\langle \bar{h}_\beta| {\bf W}_{\text{BSS}}(h_\beta) |h_\beta\rangle 
= (-\kappa)^{2S}\,.
\ee 
Now, the resonating QSL state $|1,1\rangle$ of Eq.~(\ref{eq:11}) satisfies $J_e A_v|1,1\rangle\!=\!- |1,1\rangle$, and therefore contains the combination $\frac{1}{\sqrt{2}}\left(|h_\beta\rangle\!-\!\text{sgn}(J_e)|\bar{h}_\beta\rangle\right)$. So the ground state expectation value $W_{\text{BSS}}(h_\beta)$ of the operator ${\bf W}_{\text{BSS}}(h_\beta)$ is equal to 
\be
W_{\text{BSS}}(h_\beta)\!=\!-(-\kappa)^{2S+1}\,. 
\ee
For half-integer $S$, in particular, $W_{\text{BSS}}(h_\beta)\!=\!-1$, irrespective of $\kappa$.
For the empty hexagons, such as $h_\alpha$, a well-defined flux is already fixed by the zero-point energy, as shown by BSS.~\cite{Baskaran08} Specifically, $W_{\text{BSS}}(h_\alpha)\!=\!(-1)^{\lambda S}$, where $\lambda\!=\!\kappa (\eta_1 \!+\! \eta_3 \!+\! \eta_5 )\!+\!\eta_2\!+\!\eta_4\!+\!\eta_6$, which is even. So, for integer $S$, $W_{\text{BSS}}(h_\alpha)$ is always equal to $1$, while for half-integer $S$, $W_{\text{BSS}}(h_\alpha)\!=\!-\kappa B_{h_\alpha}\!=\!-1$, because of the ABC condition on spin waves.

The BSS fluxes are in fact well defined in all eigenstates of (\ref{eq:TC1}), not just in the ground states. An excited state with an electric charge sitting on $h_\beta$ has $W_{\text{BSS}}(h_\beta)\!=\!(-\kappa)^{2S+1}$, opposite to the one in the ground state. On the other hand, an excited state with a magnetic charge on $h_\alpha$ has again $W_{\text{BSS}}(h_\alpha)\!=\!1$ for integer $S$ as in the ground state, but $+1$ for half-integer $S$. These results also mean that i) magnetic fluxes are related to BSS fluxes on empty hexagons for all $S$, and ii) electric charges are related to BSS fluxes on non-empty hexagons for half-integer $S$.

More generally, the fact that the BSS fluxes are well defined on all hexagons is consistent with Elitzur's theorem~\cite{Elitzur1975,fradkin2013field,Batista2005} that local gauge symmetries cannot be broken spontaneously. Following the works of \cite{Baskaran2007,Baskaran08}, this also necessitates that static and dynamic two-spin correlation functions are identically zero beyond NN separation, consistent with the Toric code description.

\vspace*{0.1cm}\noindent
{\bf Spin wave modes.} In the frozen dimer pattern of Fig.~\ref{fig:star&Toric}\,(a), the local Hilbert space for each spin-$S$ dimer has dimension $(2S\!+\!1)^2$, and Eq.~(\ref{eq:TC1}) describes the dynamics inside the subspace of $|m_1,m_2\rangle\!=\!|S,\kappa S\rangle$ and $|\text{-}S,\text{-}\kappa S\rangle$, where the projections $m_1$ and $m_2$ are defined along the local quantization axes. 
To this Hamiltonian (\ref{eq:TC1}), we should also add the terms that describe the coherent spin-wave bosonic modes 
\be
\mc{H}_{\text{magn}}(\{B_p\}) = \sum\nolimits_{i=1}^{N} \omega_i (\{B_p\}) ~ b_i^\dagger b_i\,,
\ee 
describing the elementary, single-particle excursions outside this $2\!\times\!2$ manifold, with $\Delta m\!=\!\pm 1$. 
Note that the important constants arising from the spin wave theory have been assigned to $J_m$ already, and that the $b_i$ bosons are the eigenmodes of the spin-wave Hamiltonian, either at the quadratic or the self-consistent quartic order [see Supplementing material, Eq. (A22)].
Also, as mentioned above, the spin-wave frequencies $\omega_i$ depend on the set $\{B_p\}$ only, and are therefore the same for all states with the same $\{B_p\}$ but different $\{A_v\}$. This entails a huge, $2^{\frac{N}{3}+2g-1}$-fold degeneracy in the spin-wave branches, for each given set of $\{B_p\}$.
We emphasize that the magnons discussed here do not describe the elementary excitations above some magnetically ordered state. Instead, they describe coherent excitations that are present in the spectrum independently of the elementary flux and charge excitations. 

\begin{figure}[!t] 
\vspace*{-0.25cm}
\includegraphics[width=0.43\textwidth,angle=0,clip=true,trim=0 0 0 0]{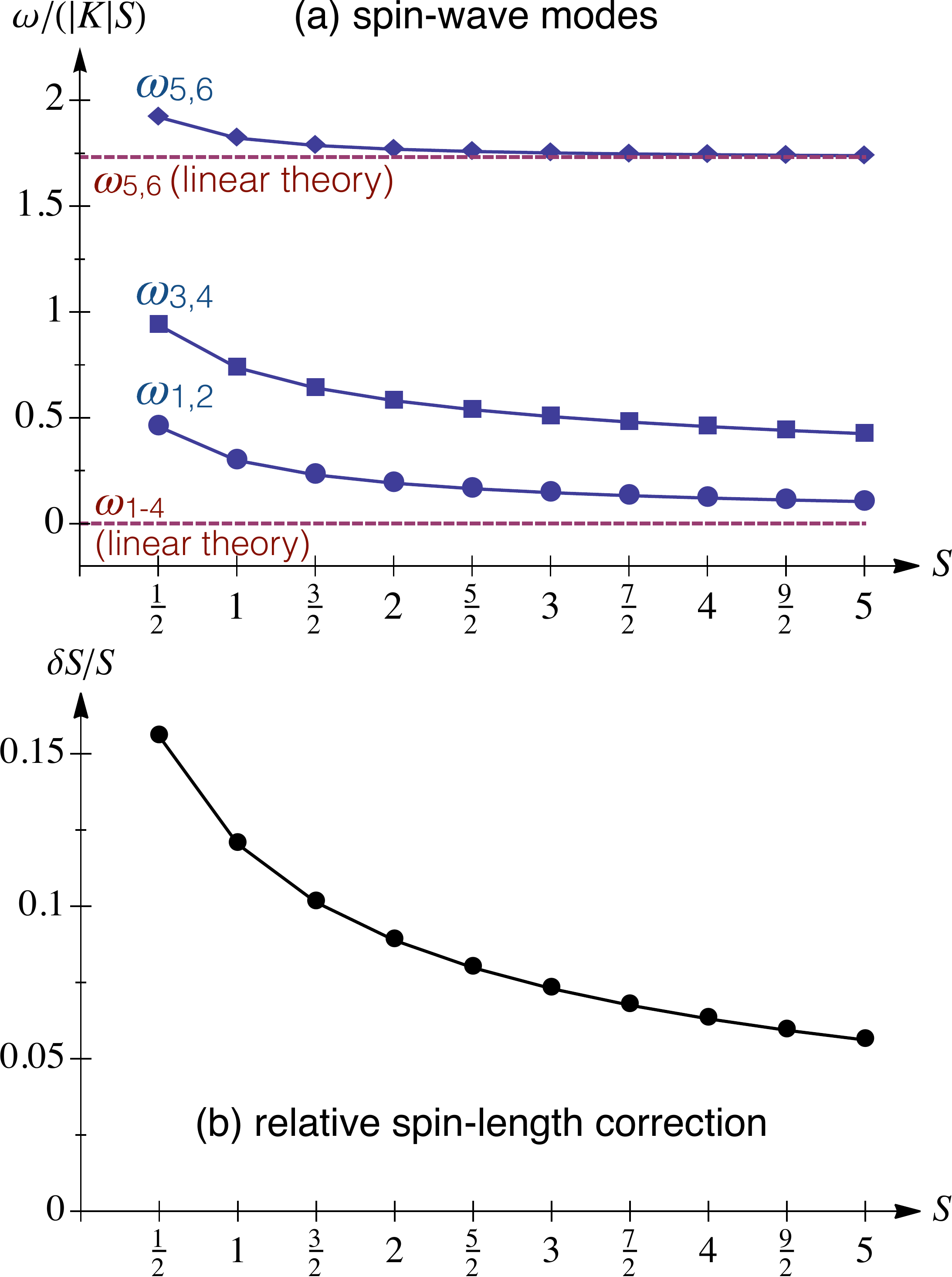}
\vspace*{-0.25cm}
\caption{(a) Magnon frequencies (in units of $|K| S$) from linear (dashed)~\cite{Baskaran08} and non-linear spin wave theory (solid), above the classical states associated with the dimer pattern of Fig.~\ref{fig:star&Toric}\,(a). (b) Relative spin length correction $\delta S/S$ from interacting spin waves.
}\label{fig:Omegas}
\end{figure}

We now examine the actual structure of the magnon spectrum. At the quadratic level, BSS have shown~\cite{Baskaran08} that the spectrum consists of six flat bands, with $\omega_{1\text{-}4}({\bf k})\!=\!0$ and $\omega_{5,6}({\bf k})\!=\!\sqrt{3}|K|S$, where the momentum ${\bf k}$ belongs to the magnetic Brillouin zone. 
However, the problem with the quadratic theory is that the modes $1$-$4$ are not true zero modes, i.e. they will be gapped out by interactions. Such spurious zero modes are typical~\cite{ChandraDoucot1988,Harris1992,Chubukov1992,Khaliullin2001,Dorier2005,Mulder2010,Chalker,francisites2015,ioannisK1K2} artifacts of the harmonic theory and reflect the modes that connect different classical minima. As commented above, the large number of such spurious zero modes in the present model leads to unreliable estimates for the relevant energy scales of the problem.
This necessitates that we push the semiclassical expansion to quartic order, and treat the problem via a standard self-consistent decoupling scheme (see Supplementing material). 

A key finding of this analysis is that spin waves remain localized inside the empty hexagons even at the interacting spin wave level, because of the local conservation laws associated with the BSS fluxes. In the language of the Holstein-Primakoff bosons, $a_i$, the conservation of BSS fluxes on empty hexagons (which remain well defined in the classical states of the `star' pattern) translates into the conservation of the parity of the total number of magnons inside the empty hexagons (see Supplementing Material). As a result, individual hopping of magnons from one empty hexagon to the next is forbidden by symmetry. Pair hopping also does not occur because, as discussed above, the inter-hexagon couplings take the form $S_i^w S_j^w$, which gives rise to a term of the type $a_i^\dagger a_i a_j^\dagger a_j$, that leaves no room for pair hopping upon decoupling.
Altogether then, the $2^{\frac{N}{3}+2g-1}$-fold degenerate branches corresponding to a given flux sector $\{B_p\}$ are flat in momentum space.

Fig.~\ref{fig:Omegas}\,(a) shows the magnon frequencies for the ground state flux sector, where all $B_p\!=\!1$, along with the corresponding results from the quadratic theory.  
All spurious modes are gapped out, and the spectrum organizes into three degenerate pairs due to symmetry (see Supplementing material). 
This figure also tells us that all modes sit far above the energy scales $|J_m|$ and $|J_e|$ of Eq.~(\ref{eq:TC1}). 
In addition, the spin length corrections $\delta S$ of Fig.~\ref{fig:Omegas}\,(b) shows that spin waves do not reduce the spin length appreciably (at maximum it is about 15\% for $S\!=\!1/2$), so the ${\bs \eta}$ variables are well defined objects.

\begin{figure}[!b]
\includegraphics[width=0.45\textwidth]{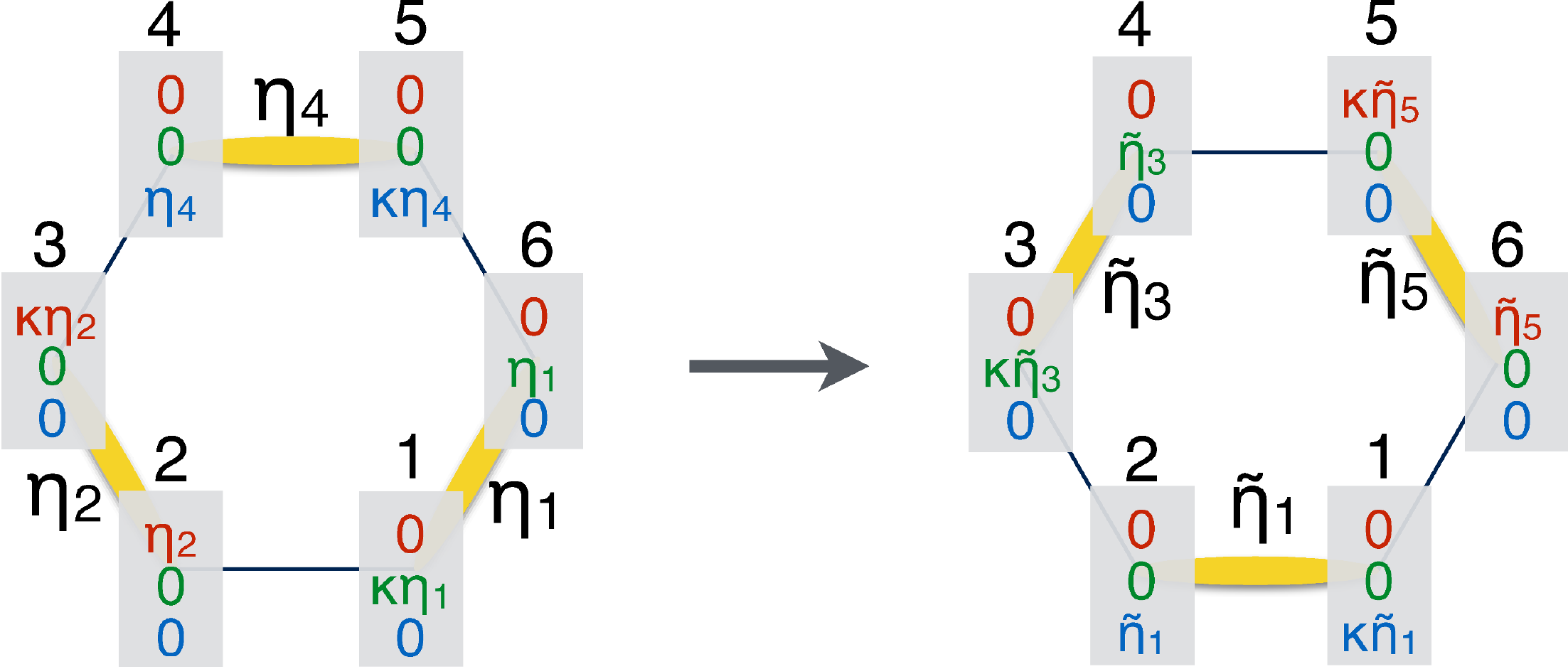}
\caption{\label{fig:DimerMoves} Tunneling process that shifts the dimers around a hexagon.} 
\end{figure}

\begin{figure}[!t] 
\vspace*{-0.25cm}
\includegraphics[width=0.23\textwidth,angle=0,clip=true,trim=0 0 0 0]{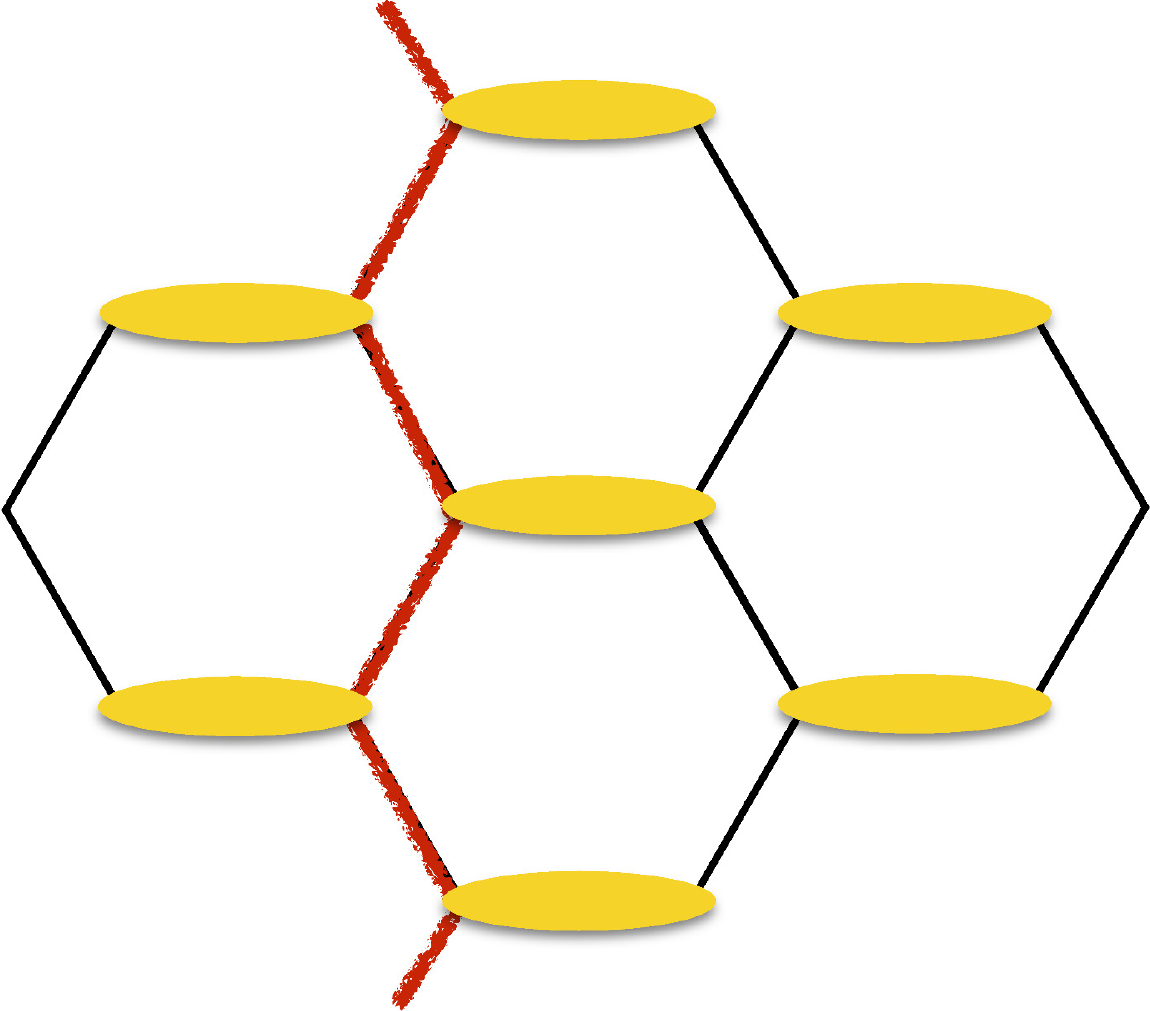}
\vspace*{-0.25cm}
\caption{Dimer pattern where empty bonds form infinite strings (red).}\label{fig:KzStates}
\end{figure}

\begin{figure}[!!t] 
\vspace*{-0.25cm}\includegraphics[width=0.4\textwidth,angle=0,clip=true,trim=0 0 0 0]{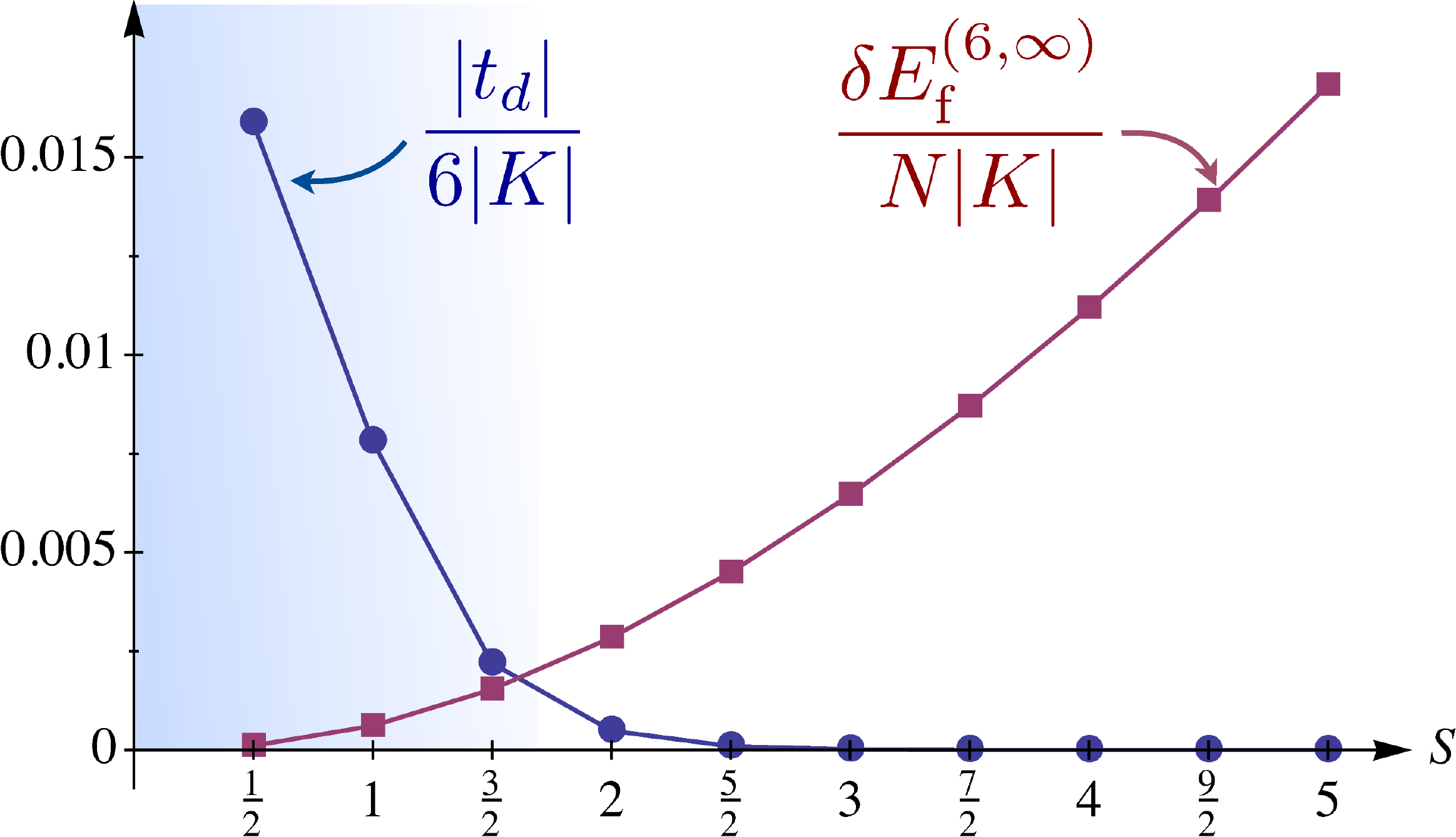}\vspace*{-0.25cm}\caption{Competition between kinetic and potential (dimer freezing) energy of dimers.}\label{fig:tD}
\end{figure}

\vspace*{0.1cm}\noindent
{\bf Physics at low $S$.} We now turn our discussion to what can go wrong with the above semiclassical picture as we lower $S$. The dimer freezing in the star pattern of Fig.~\ref{fig:star&Toric}\,(a) stems from the zero-point energy of spin waves. However, this analysis disregards the quantum tunneling between different dimer patterns. The leading process is the one around a hexagon, see Fig.~\ref{fig:DimerMoves}. The states associated with different dimer patterns are not orthonormal, but we can estimate the relevant tunneling amplitude $t_d$ using the truncation method of \cite{IoannisZ2} (see Methods):
\be\label{eq:td}
|t_d| / |K|= 3 S^2 2^{-6S} / (1-2^{-12S}).
\ee
At large $S$, $t_d$ is extremely small, and the spin-wave analysis of the dimer freezing has solid ground. This would in fact remain true down to $S\!=\!1$, if we were to use linear spin wave theory. However, this theory overestimates strongly the freezing energy scale (like $|J_m|$) due to the spurious zero modes mentioned above. As a result, $t_d$ becomes relevant below $S\!\sim\!3/2$. To see this, let us take as a representative freezing energy scale, the energy difference $\delta E_{\text{f}}^{(6,\infty)}$ between the star pattern and the `staggered' pattern of Fig.~\ref{fig:KzStates}, where the empty loops have infinite length. At the level of interacting spin wave theory, these energies are shown in Fig.~\ref{fig:tD} along with $|t_d|$ (where we divide by $N$ and by $6$, respectively, so that we compare energies per site). The results show clearly that dimers become mobile below $S\!\sim\!3/2$. (By contrast, linear spin-wave theory gives $\delta E_{\text{f}}^{(6,\infty)}\!/(NK) \!=\!(\frac{\sqrt{3}}{6}\!-\!\frac{1}{\pi})S$,~\cite{Baskaran08} which is much larger than $|t_d|/6$ down to $S\!=\!1$.)

It follows that in order to understand the physics of the $S\!=\!3/2$ and $S\!=\!1$ cases, we need to return to the Cartesian basis, and allow both the position of the dimers and their spin orientation to resonate. 
Such a `decorated quantum dimer' description may appear quite more involved, but it may actually not be the case for the particular $S\!=\!1$ case. 
The reason is that $t_d/6$ is more than ten times larger than $\delta E_{\text{f}}/N$ for $S\!=\!1$ (see Fig.~\ref{fig:tD}) and, from the standard quantum dimer model on the honeycomb lattice,~\cite{Moessner2001,Schlittler2015b} we know that $t_d$ stabilizes a resonating `plaquette' dimer pattern, known also from the context of the frustrated Heisenberg model~\cite{Albuquerque2011,Ganesh2013,Zhu2013}. Including the much smaller $J_e$ term will include the resonances with the dimers of the opposite spin orientations. 
It would be interesting to check numerically this generalized semiclassical picture for $S\!=\!1$, and moreover whether certain features of this picture carry over to the exactly solvable $S\!=\!1/2$ case.

\vspace*{0.1cm}\noindent
{\bf Discussion.}
It is shown that the low-energy sector of the large-$S$ Kitaev honeycomb model is described by a Toric code on a honeycomb superlattice. This should be contrasted with the effective square-lattice Toric code that arises in the spin-1/2 model when one of the three types of bonds has much stronger coupling than the other two.~\cite{Kitaev2006} 
Here, the magnetic and electric flux terms of the effective description arise respectively from the zero-point energy of spin waves and quantum-mechanical tunneling between different orientations of frozen dimers. This picture breaks down for $S\!\lesssim\!3/2$ where tunneling between different dimer patterns becomes relevant.

The prospects for realizing $S\!>\!1/2$ Kitaev magnets remain at present limited, although there are reports for nearly perfect honeycomb magnets with Co$^{2+}$ ions, such as Na$_2$Co$_2$TeO$_6$ and Na$_3$Co$_2$SbO$_6$,~\cite{Viciu2007} with peculiar spatial magnetic correlations.~\cite{Simonet2016} These systems show single-ion anisotropy, but it is worth checking via {\it ab initio} methods if a strong Kitaev term is also present, as in the layered spin-1/2 iridates and ruthenates.~\cite{Jackeli2009,Jackeli2010,Krempa2014,Trebst2017}
In parallel, there are proposals for emulating the model with trapped ions~\cite{Schmied2011}, superconducting quantum circuits,~\cite{You2010} coupled cavity arrays,~\cite{Xiang2012} and ultracold atoms in optical lattices,~\cite{Duan2003,Micheli2006,Gorshkov2013,Salvatore2013} which in particular offer the possibility for $S\!>\!1/2$ extensions of the model.~\cite{Micheli2006,Gorshkov2013,Salvatore2013}

Finally, we point out that the uniform~\cite{Misguich02} or staggered~\cite{Wan2013,Hwang2015} charge sectors of Eq.~(\ref{eq:TC1}) describe another well known $Z_2$ spin liquid, the RVB state of the spin-1/2 Heisenberg kagome antiferromagnet.~\cite{Sachdev92,YanHuseWhite2011,Shollwock2012,Jiang2012,IoannisZ2,ArnaudKagome} This highlights the universal topological features of QSLs arising from very different settings, across both isotropic and highly anisotropic magnets.

\vspace*{0.1cm}\noindent
{\bf Acknowledgements.} 
We thank G. Baskaran, A. Ralko, A. Tsirlin, Y. Wan and M. D. Schulz for fruitful discussions. 
Part of this work was done at the Perimeter Institute in Waterloo, which is supported by the Government of Canada through Industry Canada and by the Province of Ontario through the Ministry of Economic Development and Innovation.
We also acknowledge the support from NSF Grant No. DMR-1511768.


\small

\vspace*{0.15cm}
\noindent{\bf Methods}\\
%
\noindent{\bf Derivation of Eq.~(\ref{eq:td}).}
To calculate the tunneling $t_d$ around a single hexagon we consider the simplest $2\!\times\!2$ truncation approach described in \cite{IoannisZ2} (see also \cite{Schwandt2010}). Namely, we take a hexagon cluster and project the Hamiltonian into the $2\!\times\!2$ basis of dimer states shown in Fig.~\ref{fig:DimerMoves}:
\bea
\begin{array}{c}
|1\rangle = |\kappa \eta_1{\bf y}\rangle_1 ~|\eta_2{\bf x}\rangle_2 ~|\kappa\eta_2{\bf x}\rangle_3 ~|\eta_4{\bf z}\rangle_4 ~|\kappa\eta_4{\bf z}\rangle_5 ~|\eta_1{\bf y}\rangle_6\\
|2\rangle = |\kappa \tilde{\eta}_1{\bf z}\rangle_1 ~|\tilde{\eta}_1{\bf z}\rangle_2 ~|\kappa \tilde{\eta}_3{\bf y}\rangle_3 ~|\tilde{\eta}_3{\bf y}\rangle_4 ~|\kappa \tilde{\eta}_5{\bf x}\rangle_5 ~|\tilde{\eta}_5{\bf x}\rangle_6
\end{array}\,.
\eea
The magnitude of the overlap $\Omega$ between the two states is
\bea
|\Omega| = |\langle 1|2\rangle| = 2^{-6S}\,,
\eea
and the matrix elements of the cluster Hamiltonian are
\small
\bea
\begin{array}{c}
E_0 \equiv \langle 1|\mc{H}|1\rangle= \langle 2|\mc{H}|2\rangle = -3|K| S^2 \\
v\equiv \langle 1|\mc{H}|2\rangle=-6|K| S^2\Omega
\end{array}
\eea
Orthonormalizing the basis leads to the effective $2\!\times\!2$ Hamiltonian $\left(\!\!\!\begin{array}{cc}
E_0\!+\!v \!\!&\!\! t_d \\
t_d \!\!&\!\! E_0\!+\!v 
\end{array}\!\!\!\right)$, where the tunneling amplitude $t_d$ and the potential energy $V$ are given by~\cite{IoannisZ2,Schwandt2010}
\bea
t_d = \frac{v-E_0\Omega}{1-\Omega^2} 
= -\frac{3KS^2 2^{-6S}}{1-2^{-12S}}\times\text{sgn}(\Omega),~~
V=-\Omega t_d\,.
\eea
The latter is much smaller than $|t_d|$ and can be ignored.


\vspace*{0.1cm}
\noindent{\bf Author contributions}\\ 
All authors contributed to the analysis and interpretation of the results, and the preparation of the manuscript.

\vspace*{0.1cm}
\noindent{\bf Competing financial interests}\\ 
The authors declare no competing financial interests.

\normalsize

%

\clearpage

\appendix

\pagenumbering{roman}
\widetext

\begin{widetext}

\begin{center}
{\Large {\bf Supplemental material}}
\end{center}





\section{Semiclassical expansion around the states associated with the star dimer pattern}
\vspace*{-0.25cm}
\subsection{Lattice superstructure \& Hamiltonian}
\vspace*{-0.25cm}
Here we provide the details of the semiclassical expansion around the  states of the star dimer pattern of Fig.~2 of the main text. 
To this end, we shall use the six-sublattice decomposition of Fig.~\ref{fig:ABC}, with a superlattice defined by the primitive translation vectors ${\bf T}_1$ and ${\bf T}_2$. Any given site $i$ of the lattice can be labeled as $i=(\vec{R},\nu)$, where ${\bf R}$ is a primitive vector of the superlattice and $\nu=1$-$6$ is the sublattice index. In this parametrization, the positions of the empty hexagons $h_\alpha$ are labeled by ${\bf R}$. 
The classical state is parametrized in terms of the $\eta$-variables,  as shown in Fig.~2 of the main text.
We will also use the local coordinate frames given in Eq.~(3) of the main text, and define for each empty hexagon $h_{{\bf R}}$
\small
\be
\gamma_{{\bf R}} \equiv \kappa B_{\bf R}\,.
\ee 
\normalsize
With these conventions and definitions, the Hamiltonian reads
\small
\bea
\mc{H}&=&
K \sum_{\vec{R}}\big[
S_{\vec{R},1}^uS_{\vec{R},2}^u+S_{\vec{R},2}^vS_{\vec{R},3}^v+S_{\vec{R},3}^uS_{\vec{R},4}^u+S_{\vec{R},4}^vS_{\vec{R},5}^v+S_{\vec{R},5}^uS_{\vec{R},6}^u-\gamma_{\bf R} S_{\vec{R},6}^v S_{\vec{R},1}^v \big]
\nonumber\\&&
-|K| \sum_{\vec{R}} \big[S_{\vec{R},3}^wS_{\vec{R}-\vec{T}_1,6}^w + S_{\vec{R},1}^wS_{\vec{R}+{\bf T}_1-{\bf T}_2,4}^w +  S_{\vec{R},5}^wS_{\vec{R}+\vec{T}_2,2}^w
\big]\,.
\eea
\normalsize

\begin{figure}[!b]
\centering
\includegraphics[width=0.3\textwidth]{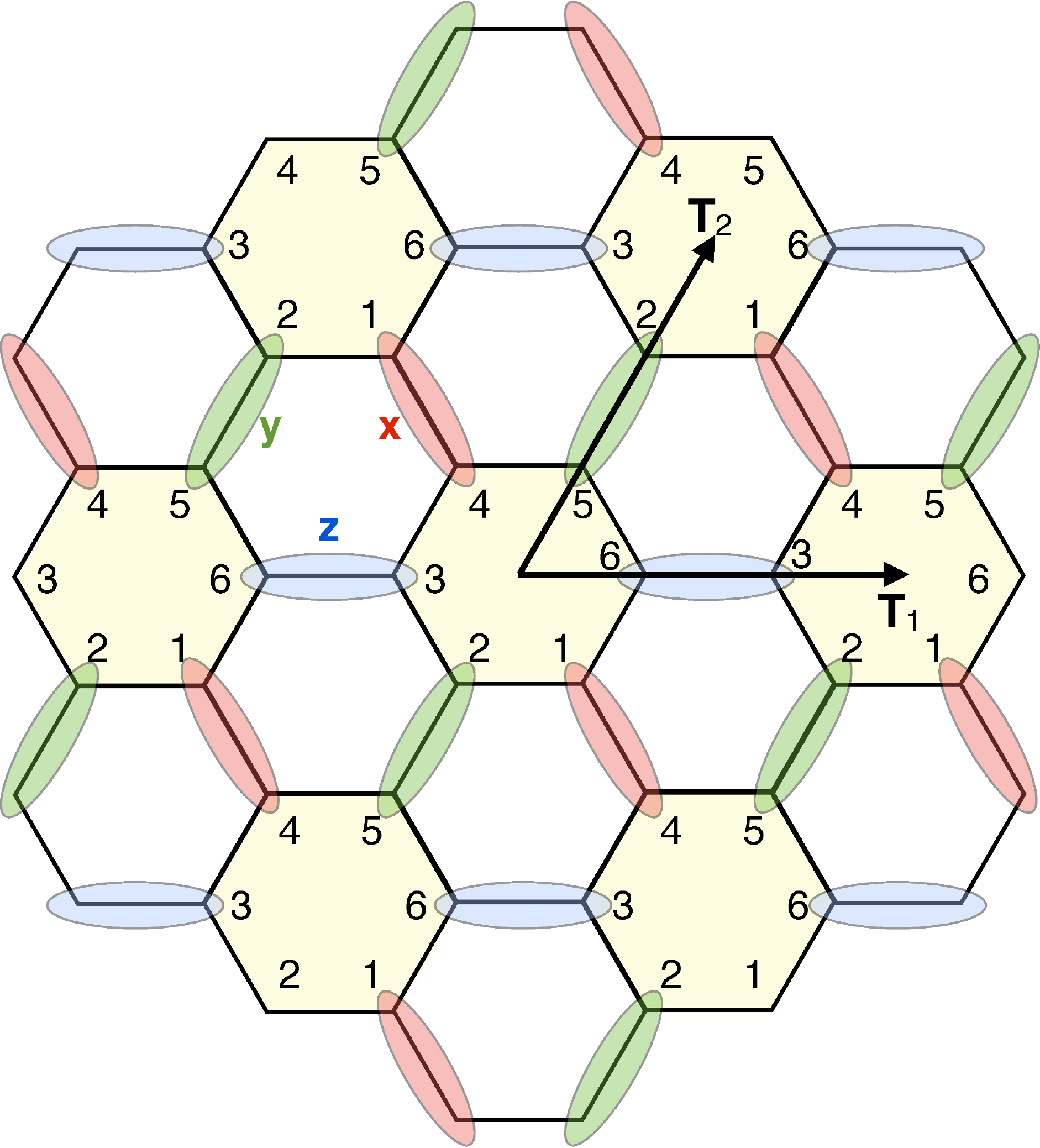}
\caption{\label{fig:ABC} The `star' dimer pattern of Fig.~2 of the main text and the six-sublattice decomposition used here.} 
\end{figure}

\vspace*{-0.75cm}

\vspace*{-0.25cm}
\subsection{Semiclassical expansion}
\vspace*{-0.25cm}
In our semiclassical expansion we will keep up to four boson terms. So it suffices to keep the following terms from the standard~\cite{HP1940SM} Holstein-Primakoff expansion for each site $i=({\bf R},\nu)$:
\small
\bea
&& S_{i}^w = S-c_i^+ c_i,~~S_{i}^+ \simeq \sqrt{2S}(c_i -\frac{n_i}{4 S} c_i),~~
S_{i}^- \simeq \sqrt{2S}(c_i^+ -c_i^+\frac{n_i}{4 S} )\\
&&S_{i}^u \simeq \frac{\sqrt{S}}{\sqrt{2}} (c_i + c_i^+ - \frac{n_i}{4 S}  c_i - c_i^+ \frac{n_i}{4 S}  ),~~
S_{i}^v \simeq -i \frac{\sqrt{S}}{\sqrt{2}} (c_i - c_i^+ - \frac{n_i}{4 S}  c_i + c_i^+ \frac{n_i}{4 S}  ) 
\eea
\normalsize
where $c_i, c_i^+$ are bosonic operators. We have:
\small
\bea
&&
S_i^u S_j^u \simeq \frac{S}{2}\Big(c_i c_j + c_i c_j^+ + h.c.\Big) -\frac{1}{8}\Big(
c_i n_j c_j +c_i^+ n_j c_j + c_j n_i c_i + c_j^+ n_i c_i + h.c.\Big)\nonumber\\
&&
S_i^v S_j^v \simeq -\frac{S}{2}\Big(c_i c_j - c_i c_j^+ + h.c.\Big) + \frac{1}{8}\Big(
c_i n_j c_j - c_i^+ n_j c_j + c_j n_i c_i - c_j^+ n_i c_i + h.c.\Big)\nonumber
\eea
\normalsize
Below we shall make use of the following mean-field parameters 
\small
\be
\boxed{p_i=\langle c^+_i c_i \rangle},~~~
\boxed{q_i=\langle c_i c_i \rangle},~~~
\boxed{m_{ij}=\langle c_i c_j^+\rangle},~~~ 
\boxed{\delta_{ij}=\langle c_i c_j\rangle}\,.
\ee
\normalsize
These parameters are all real numbers because when written in the local coordinate frames above, the Hamiltonian has real matrix elements, and in addition the states around which we expand are real. 
This also implies the relations $m_{ij}=m_{ji}$ and $\delta_{ij}=\delta_{ji}$.
Next, we can decouple the quartic terms as follows:
\small
\bea
&&
c_i n_j c_j \simeq m_{ij} c_j c_j +2 \delta_{ij} n_j +2 p_j c_i c_j +q_j c_i c_j^+-m_{ij} q_j -2\delta_{ij} p_j \nonumber\\
&&
c_i^+ n_j c_j \simeq \delta_{ij} c_j c_j +2 m_{ij} n_j +2 p_j c_i^+ c_j +q_j c_i^+ c_j^+-\delta_{ij} q_j -2m_{ij} p_j \nonumber
\eea
\normalsize

Let us now write down the resulting expressions for each type of interaction that appears in the Hamiltonian.
\begin{itemize}
\item 
{\bf Terms of the type $S_{i}^uS_{j}^u$ (where $i$ and $j$ belong to the same empty hexagon):}
\small
\bea
\boxed{
S_i^u S_j^u \simeq 
\tau_{ij}
+\Big(
f_{ji} c_j c_j + f_{ij} c_i c_i 
+ g_{ij} c_i c_j
+ g_{ij} c_i c_j^+
+ h.c.\Big)
+4 f_{ij} (n_j +n_i)
}\label{eq:SiuSju}
\eea
\normalsize 
where 
\small
\be\label{eq:fgtau}
\boxed{
f_{ij}=-\frac{m_{ij}+\delta_{ij}}{8},~~~
g_{ij}=\frac{S}{2}-\frac{2 (p_i+p_j)+q_i+q_j}{8},~~~
\tau_{ij}=-2 f_{ij} [ q_i+q_j+2 (p_i+p_j)]
}
\ee
\normalsize

\item
{\bf Terms of the type $S_{i}^vS_{j}^v$ (where $i$ and $j$ belong to the same empty hexagon):}
\small
\bea
\boxed{
S_i^v S_j^v \simeq 
\tau'_{ij}
+ \Big(
{f_{ij}'} c_j c_j
+{f_{ij}'} c_i c_i 
+g'_{ij} c_i c_j
-g'_{ij}  c_i c_j^+
+ h.c.
\Big)
-4 f'_{ij}  (n_i+n_j)}  
\label{eq:SivSjv}
\eea
\normalsize 
where 
\small
\be\label{eq:fpgptaup}
\boxed{
f'_{ij}\!=\!\frac{m_{ij}\!-\!\delta_{ij}}{8},~~~ 
g'_{ij}\!=\!-\frac{S}{2}\!+\!\frac{2 (p_i+p_j)-q_i-q_j}{8},~~~ 
\tau'_{ij}\!=\!-2f_{ij}' [q_i+q_j-2 (p_i+p_j)]
} 
\ee
\normalsize 

\item
{\bf Terms of the type $S_{{\bf R},\nu}^wS_{{\bf R}+{\bf T}_{\nu\mu},\mu}^w$:} The terms that couple different empty hexagons are of the form $S_{{\bf R},\nu}^wS_{{\bf R}+{\bf T}_{\nu\mu},\mu}^w$. For simplicity, we will label $({\bf R},\nu)\to i$ and $({\bf R}+{\bf T}_{\nu\mu},\mu)\to j$. We have:
\small
\bea
S_i^wS_j^w = \big(S-n_i\big)\big(S-n_j\big) =  \big( S^2 -S (n_i+n_j) + n_i n_j \big)\,.
\eea
\normalsize
The quartic term decouples as follows:
\small
\bea
n_i n_j &\simeq&
\big( p_j n_i + p_i n_j \big) +  \big( \delta_{ij} c_i c_j +h.c.\big)
+ \big( m_{ij} c_i c^+_j+ h.c  \big)-p_i p_j-\delta_{ij}^2-m_{ij}^2\,.
\eea
\normalsize
Now, the state around which we expand does not break the local BSS flux operators defined on the empty hexagons: 
\small
\bea
W_{\text{BSS}}({\bf R}) &=& \exp\!\big\{i \pi \big[ S_{\bf R,1}^x+S_{\bf R,2}^y+S_{\bf R,3}^z+S_{\bf R,4}^x+S_{\bf R,5}^y+S_{\bf R,6}^z \big] \big\} \nonumber\\
&=& \exp\!\big\{i \pi \big[ \kappa \big( \eta_1 S_{\bf R,1}^w+\eta_3 S_{\bf R,3}^w+\eta_5 S_{\bf R,5}^w\big)+\big(\eta_2 S_{\bf R,2}^w+\eta_4 S_{\bf R,4}^w+\eta_6 S_{\bf R,6}^w\big) \big] \big\} \nonumber\\
&=& (-1)^{\lambda_{\bf R} S} 
\exp\!\big\{-i \pi \big[ \kappa \big( \eta_1 n_{\bf R,1}+\eta_3 n_{\bf R,3}+\eta_5 n_{\bf R,5}\big)+\big(\eta_2 n_{\bf R,2}+\eta_4 n_{\bf R,4}+\eta_6 n_{\bf R,6}\big) \big] \big\}\,,\label{eq:WR}
\eea
\normalsize
where $n_i = c_i^+ c_i$ is the boson number operator and $\lambda_{\bf R}=\kappa(\eta_{{\bf R},1}+\eta_{{\bf R},3}+\eta_{{\bf R},5})+(\eta_{{\bf R},2}+\eta_{{\bf R},4}+\eta_{{\bf R},6})$, see main text.
The invariance of the Hamiltonian and the state around which we expand under this operation translates into the invariance of the parity of the number 
$\kappa\big(\eta_1 n_{\bf R,1}+\eta_3 n_{\bf R,3}+\eta_5 n_{\bf R,5}\big)+\big(\eta_2 n_{\bf R,2}+\eta_4 n_{\bf R,4}+\eta_6 n_{\bf R,6}\big)$. But since $\kappa$ and $\eta$ can only take the values $+1$ and $-1$, it follows that the parity of this number is the same as the parity of the total number $N_{\bf R}$ of bosons in any given empty hexagon:
\small
\be
N_{\bf R} = \sum_{\nu=1-6} n_{\bf R,\nu}\,.
\ee
\normalsize
This means that terms that change the parity of $N_{\bf R}$ are not allowed in the expansion. This excludes terms of the type $c_i c_j$ or $c_i c^+_j$, where $i$ and $j$ belong to different empty hexagons (see definition above). Equivalently, the mean-field parameters $m_{ij}$ and $\delta_{ij}$ vanish by symmetry, and this is true to all orders in the Holstein-Primakoff expansion.
We therefore get:
\small
\bea
&&
 S_i^w S_j^w \simeq 
\Big\{ \frac{S^2-p_i p_j}{2}+(p_j-S) n_i \Big\}
+ \Big\{ \frac{S^2-p_i p_j}{2}+(p_i-S) n_j \Big\}\,,
\nonumber
\eea
\normalsize
i.e. empty hexagons decouple from each other and the $S_i^w S_j^w$ terms give, for each empty hexagon ${\bf R}$ alone, a contribution
\small
\bea
\boxed{\frac{S^2-p_{i} p_{j}}{2}+(p_{j}-S) n_{i}}
\eea
\normalsize 
where the constant $p_{j}=p_{{\bf R}+{\bf T}_{\nu\mu},\mu}$ refers to a neighboring hexagon and has to be found self-consistently in the general case. 

\end{itemize}

It is useful to add here one more consequence of the BSS flux conservation. In the classical, reference state, where all $n_i$ vanish, the BSS fluxes are equal to $(-1)^{\lambda_{\bf R} S}$ (see main text and \cite{Baskaran08SM}). Spin wave fluctuations dress the reference state but cannot change the BSS fluxes, because these are integer numbers. Equation~(\ref{eq:WR})  then implies that the dressed ground state contains only terms with an even number of bosons $N_{\bf R}$.

\vspace*{-0.25cm}
\subsection{Semiclassical expansion around the states of the star dimer pattern with uniform $\gamma_{\bf R}$}
\vspace*{-0.25cm}
In the following we shall focus on the classical states that have the same $\gamma_{\bf R}$ on all empty hexagons. This means that $p_{{\bf R},\nu}$ is independent of ${\bf R}$, and we can therefore replace $p_{{\bf R}+{\bf T}_{\nu\mu},\mu}\to p_\mu$ in the above contribution from the $S_i^w S_j^w$  terms.
Collecting all terms referring to a given hexagon and dropping the index ${\bf R}$ we get:
\small
\bea
\mc{H}/|K|&=& f_0 + \frac{1}{2}\sum_\nu\Big\{
d_\nu c^+_\nu c_\nu 
+ d_\nu' c_\nu c_\nu
+  \lambda_{\nu,\nu+1} c_\nu c_{\nu+1}
+ \lambda_{\nu+1,\nu} c_{\nu+1} c_\nu
 +\pi_{\nu,\nu+1} c_\nu c^+_{\nu+1} + h.c. 
 \Big\}\,,
\eea
\normalsize
where we have defined:
\small
\bea
&&
f_0 = 
- 3 S^2
+ \big( p_1 p_4 +p_2 p_5 + p_3 p_6 \big) 
-\kappa \big( \tau_{12}+\tau_{34}+\tau_{56}+\tau'_{23}+\tau'_{45}-\gamma\tau'_{61}\big)
-\frac{1}{2}\sum_\nu d_\nu\,,\nonumber\\
&&
d_1 = S-p_4 - 4 \kappa~(f_{12}+\gamma f_{61}'),~~~ d_1' = -2 \kappa (f_{12}-\gamma {f_{16}'}),~~~
\lambda_{12} = -\kappa g_{12},~~~~ \pi_{12} = \lambda_{12} \nonumber\\
&&
d_2 = S-p_5 - 4 \kappa~(f_{12}-f_{23}'),~~~ d_2' = -2\kappa (f_{21}+{f_{23}'}),~~~
\lambda_{23}=-\kappa g'_{23},~~~~ \pi_{23} = -\lambda_{23} \nonumber\\
&&
d_3 = S-p_6 - 4 \kappa~(f_{34}-f_{23}'),~~~ d_3' = -2\kappa (f_{34}+{f_{32}'}),~~~
\lambda_{34} = -\kappa g_{34},~~~~ \pi_{34} = \lambda_{34} \nonumber\\
&&
d_4 = S-p_1 - 4 \kappa~(f_{34}-f_{45}'),~~~ d_4' = -2\kappa (f_{43}+{f_{45}'}),~~~
\lambda_{45}=-\kappa g'_{45} ,~~~~ \pi_{45} = -\lambda_{45}\nonumber\\
&&
d_5 = S-p_2 - 4 \kappa~(f_{56}-f_{45}'),~~~ d_5' = -2\kappa (f_{56}+{f_{54}'}),~~~
\lambda_{56}=-\kappa g_{56} ,~~~~ \pi_{56} = \lambda_{56}\nonumber\\
&&
d_6 = S-p_3 - 4 \kappa~(f_{56}+\gamma f_{61}'),~~~ d_6' = -2\kappa (f_{65}-\gamma {f_{61}'}),~~~
\lambda_{61}=\gamma\kappa g'_{61} ,~~~~ \pi_{61} = -\lambda_{61}~.\nonumber
\eea
\normalsize
Next we define $\vec{C}^+=\left(c_{1}^+, \cdots c_{6}^+, c_{1}, \cdots c_{6} \right)$ and write:
\small
\bea
\mc{H}&=& f_0  + \frac{1}{2}  \vec{C}^+ \cdot \vec{M} \cdot \vec{C}\,,
\eea
\normalsize
where the nonzero matrix elements of the matrix $\vec{M}$ are as follows:
\small
\bea
\vec{M}=
\left(\begin{array}{c|c|c|c|c|c V{2}  c|c|c|c|c|c}
d_1&\pi_{12}&&&&\pi_{61}&{d_1'}&\lambda_{12}&&&&\lambda_{61} \\ 
\hline
\pi_{12}&d_2&\pi_{23}&&&&\lambda_{12}&{d_2'}&\lambda_{23}&&& \\ 
\hline
&\pi_{23}&d_3&\pi_{34}&&&&\lambda_{23}&{d_3'}&\lambda_{34}&& \\
\hline
&&\pi_{34}&d_4&\pi_{45}&&&&\lambda_{34}&{d_4'}&\lambda_{45}& \\
\hline
&&&\pi_{45}&d_5&\pi_{56}&&&&\lambda_{45}&{d_5'}& \lambda_{56}\\
\hline
\pi_{61}&&&&\pi_{56}&d_6&\lambda_{61}&&&&\lambda_{56}& {d_6'}\\
\hlineB{2}
d_1'&\lambda_{12}&&&&\lambda_{61}&d_1&\pi_{12}&&&&\pi_{61}\\
\hline
\lambda_{12}&d_2'&\lambda_{23}&&&&\pi_{12}&d_2&\pi_{23}&&& \\
\hline
&\lambda_{23}&d_3'&\lambda_{34}&&&&\pi_{23}&d_3&\pi_{34}&& \\
\hline
&&\lambda_{34}&d_4'&\lambda_{45}&&&&\pi_{34}&d_4&\pi_{45}& \\
\hline
&&&\lambda_{45}&d_5'&\lambda_{56}&&&&\pi_{45}&d_5&\pi_{56} \\
\hline
\lambda_{61}&&&&\lambda_{56}&d_6'&\pi_{61}&&&&\pi_{56}& d_6\\
\end{array}
\right)
\eea
\normalsize
We next define the commutator matrix
\small
\bea
\vec{g} \!=\! \vec{C}\!\cdot\!\vec{C}^\dagger - \left( (\vec{C}^\dagger)^T \!\cdot\! \vec{C}^T \right)^T
=\left(
\begin{array}{c|c}
1\!\text{l}_6 & 0 \\
\hline
0 & -1\!\text{l}_6
\end{array}
\right)\,,
\eea
\normalsize  
where $1\!\text{l}_6$ stands for the identity $6\!\times\!6$ matrix, and then perform a standard~\cite{Blaizot} Bogoliubov transformation $\vec{C}\!=\!\vec{S}\!\cdot\!\vec{B}$, which must conserve the commutation relations $\tilde{\vec{g}}\!=\!\vec{g}$. This relation gives 
\small
\be
\vec{S}^\dagger\cdot\vec{g}\cdot\vec{S}=\vec{g} \Rightarrow 
\vec{S}^{-1}=\vec{g}\cdot\vec{S}^\dagger\cdot\vec{g}\,.
\ee
\normalsize
The matrix $\vec{S}$ must also satisfy the relation 
\small
\be\label{eq:LSL}
\bs{L} \vec{S} \bs{L}=\vec{S}^\ast,~~~\text{where}~~~
\bs{L}=\left(
\begin{array}{cc} 
0 & 1\!\text{l}_6 \\
1\!\text{l}_6 & 0
\end{array}
\right)\,,
\ee
\normalsize
and at the same time diagonalize the Hamiltonian:
\small
\bea
\mc{H}  \!=\! f_0  \!+\! \frac{1}{2} \vec{B}^\dagger \!\cdot\! (\vec{S}^\dagger \vec{M} \vec{S}) \!\cdot\! \vec{B}
=f_0\!+\! \frac{1}{2} \vec{B}^\dagger \!\cdot\! \vec{\Omega}_{M} \!\cdot\! \vec{B},
\eea
\normalsize
where $\vec{\Omega}_{M}$ is diagonal and can be found from the eigenvalue equation $(\vec{g} \vec{M}) \cdot \vec{S} = \vec{S} \cdot (\vec{g} \vec{\Omega}_{M}) \equiv \vec{S}\cdot \vec{\Omega}_{g M}$. 
It can be shown~\cite{Blaizot} that the eigenvalues of $\vec{g}\cdot\vec{M}$ come in pairs $(\omega_\nu,-\omega_\nu)$, where $\nu=1$-$6$. We finally get
\small
\be\label{eq:finaldiagform1}
\mc{H}= f_0 +  \sum_{\nu=1-6} \omega_\nu\left( b_\nu^+ b_\nu +\frac{1}{2}\right)~.
\ee
\normalsize
The ground state energy is, in particular, given by $E_0=f_0+\frac{1}{2}\sum_\nu \omega_\nu$.

\vspace*{-0.25cm}
\subsection{Mean field parameters: General relations}
\vspace*{-0.25cm}
Let us define the six eigenvectors of the matrix ${\bf g}\cdot{\bf M}$ that correspond to non-negative eigenvalues by ${\bf X}_\nu$, $\nu=1$-$6$. 
Using:
\small
\bea
c_{i} = \sum_{j=1-6} \Big( S_{i,j} b_{j}+ S_{i,6+j} b_{j}^+\Big) ,~~~~
c_{i}^+ = \sum_{j=1-6}  \Big( S_{6+i,j} b_{j}+ S_{6+i,6+j} b_{j}^+\Big)= \sum_{j=1-6}  \Big(S_{i,j+6}^\ast b_{j}+ S_{i,j}^\ast b_{j}^+\Big)
\eea
\normalsize
we get the following expressions for the mean-field parameters:
\small
\bea
&&
n_i=\langle c_{i}^+ c_{i}\rangle = \sum_{\nu=1-6} |S_{i,6+\nu}|^2 = \sum_{\nu} |S_{i+6,\nu}|^2 =  \sum_{\nu} |X_\nu(i+6)|^2\,,\nonumber\\
&&
\delta_{ij}=\langle c_{i} c_{j}\rangle = \sum_{\nu=1-6} S_{i,\nu} S_{j,6+\nu}= \sum_{\nu} S_{i,\nu} S^\ast_{j+6,\nu} 
= \sum_{\nu} X_\nu(i) X^\ast_\nu(j+6)\,,\\
&&
m_{ij}=\langle c_{i} c_{j}^+\rangle =  \sum_{\nu=1-6} S_{i,\nu} S_{j+6,\nu+6} =  
\sum_{\nu} S_{i,\nu} S_{j,\nu}^\ast = \sum_{\nu} X_\nu(i) X_\nu^\ast(j)\,, \nonumber
\eea
\normalsize
where ${\bf X}_\nu$ denotes the $\nu$-th eigenvector of ${\bf g}\cdot{\bf M}$. Note that the last expressions in each line do not depend on the arbitrary phase for the eigenvectors ${\bf X}_\nu$, which come out arbitrary when we diagonalize the matrix ${\bf g}\cdot{\bf M}$ numerically. 

\vspace*{-0.25cm}
\subsection{Mean field parameters: Symmetry constraints}
\vspace*{-0.25cm}
We have already mentioned that all mean-field parameters defined above are real quantities. 
Here we give a list of symmetry operations (of the Hamiltonian and of the classical state around which we expand) which reduce strongly the number of independent mean-field parameters.

\begin{itemize}
\item {\bf Symmetry $\bs{\Sigma}_1$}. 
This is a $\pi$-rotation in real space around the center of the hexagon, followed by $\pi/2$-rotations around the local $\bf{w}$-axes in spin space:
\small
\bea
&&
(S_1^u,S_1^v,S_1^w) \to ( S_4^v,- S_4^u,S_4^w),~~~
(S_2^u,S_2^v,S_2^w) \to ( S_5^v,- S_5^u,S_5^w),~~~
(S_3^u,S_3^v,S_3^w) \to ( S_6^v,- S_6^u,S_6^w),\nonumber\\
&&
(S_4^u,S_4^v,S_4^w) \to(-\gamma S_1^v,\gamma S_1^u,S_1^w),~~~
(S_5^u,S_5^v,S_5^w) \to (-\gamma S_2^v,\gamma S_2^u,S_2^w),~~~
(S_6^u,S_6^v,S_6^w) \to (-\gamma S_3^v,\gamma S_3^u,S_3^w)\nonumber
\eea
\normalsize
These relations are equivalent with 
\small
$
\boxed{c_1 \to -i  c_4,~~ c_2 \to -i  c_5,~~c_3 \to -i  c_6,~~c_4 \to +i \gamma c_1,~~c_5 \to  +i \gamma c_2,~~c_6 \to +i \gamma c_3}
$
\normalsize

\item {\bf Symmetry $\bs{\Sigma}_2$}. 
This is a reflection through the bonds (3,6) in real space, followed by $\pi/2$-rotations around the local $\bf{w}$-axes in spin space:
\small
\bea
&&
(S_1^u,S_1^v,S_1^w) \to (\gamma S_5^v,-\gamma S_5^u,S_5^w),~~~
(S_2^u,S_2^v,S_2^w) \to (\gamma S_4^v,-\gamma S_4^u,S_4^w),~~~
(S_3^u,S_3^v,S_3^w) \to (\gamma S_3^v,-\gamma S_3^u,S_3^w),\nonumber\\
&&
(S_4^u,S_4^v,S_4^w) \to(\gamma S_2^v,-\gamma S_2^u,S_2^w),~~~
(S_5^u,S_5^v,S_5^w) \to (\gamma S_1^v,-\gamma S_1^u,S_1^w),~~~
(S_6^u,S_6^v,S_6^w) \to (- S_6^v, S_6^u,S_6^w)\nonumber
\eea
\normalsize
These relations are equivalent with:
\small
$
\boxed{c_1 \to -i \gamma c_5,~~c_2 \to -i \gamma c_4,~~c_3 \to -i \gamma c_3,~~c_4 \to -i \gamma c_2,~~c_5 \to  -i \gamma c_1,~~c_6 \to +i  c_6}
$
\normalsize

\item {\bf Symmetry $\bs{\Sigma}_3$}.
This is a reflection through the middle of the bonds (1,2) and (4,5) in real space, followed by zero or $\pi$-rotations around the local-$\bf{w}$ axes in spin space:
\small
\bea
&&
(S_1^u,S_1^v,S_1^w) \to (-\gamma S_2^u,-\gamma S_2^v,S_2^w),~~~
(S_2^u,S_2^v,S_2^w) \to (-\gamma S_1^u,-\gamma S_1^v,S_1^w),~~~
(S_3^u,S_3^v,S_3^w) \to (S_6^u,S_6^v,S_6^w),\nonumber\\
&&
(S_4^u,S_4^v,S_4^w) \to(S_5^u,S_5^v,S_5^w),~~~
(S_5^u,S_5^v,S_5^w) \to (S_4^u,S_4^v,S_4^w),~~~
(S_6^u,S_6^v,S_6^w) \to (S_3^u,S_3^v,S_3^w)\nonumber
\eea
\normalsize
These relations are equivalent with:
\small
$
\boxed{c_1 \to -\gamma c_2,~~c_2 \to -\gamma c_1,~~c_3 \to c_6,~~c_4 \to c_5,~~c_5 \to  c_4,~~c_6 \to c_3}
$
\normalsize

\item
{\bf Symmetry $\bs{\Sigma}_4$}.
This is a reflection through the bonds (1,4) in real space, followed by $\pi/2$-rotations around the local $\bf{w}$-axes in spin space:
\small
\bea
&&
(S_1^u,S_1^v,S_1^w) \to (-\gamma S_1^v,\gamma S_1^u,S_1^w),~~~
(S_2^u,S_2^v,S_2^w) \to (S_6^v,-S_6^u,S_6^w),~~~
(S_3^u,S_3^v,S_3^w) \to (S_5^v,-S_5^u,S_5^w),\nonumber\\
&&
(S_4^u,S_4^v,S_4^w) \to(S_4^v,-S_4^u,S_4^w),~~~
(S_5^u,S_5^v,S_5^w) \to (S_3^v,-S_3^u,S_3^w),~~~
(S_6^u,S_6^v,S_6^w) \to (S_2^v,- S_2^u,S_2^w)\nonumber
\eea
\normalsize
These relations are equivalent with:
\small
$
\boxed{c_1 \to +i \gamma c_1,~~c_2 \to -i c_6,~~c_3 \to -i c_5,~~c_4 \to -i c_4,~~c_5 \to  -i c_3,~~c_6 \to -i c_2}
$
\normalsize

\item
{\bf Symmetry $\bs{\Sigma}_5$}. This is a $\pi/6$-rotation in real space, followed by a $\pi/2$-rotation around the local ${\bf w}$-axes in spin space:
\small
\bea
&&
(S_1^u,S_1^v,S_1^w) \to (S_2^v,-S_2^u,S_2^w),~~~
(S_2^u,S_2^v,S_2^w) \to (S_3^v,-S_3^u,S_3^w),~~~
(S_3^u,S_3^v,S_3^w) \to (S_4^v,-S_4^u,S_4^w),\nonumber\\
&&
(S_4^u,S_4^v,S_4^w) \to(S_5^v,-S_5^u,S_5^w),~~~
(S_5^u,S_5^v,S_5^w) \to (S_6^v,-S_6^u,S_6^w),~~~
(S_6^u,S_6^v,S_6^w) \to (-\gamma S_1^v,\gamma S_1^u,S_1^w)\nonumber
\eea
\normalsize
These relations are equivalent with:
\small
$
\boxed{c_1 \to -i c_2,~~c_2 \to -i c_3,~~c_3 \to -i c_4,~~c_4 \to -i c_5,~~c_5 \to  -i c_6,~~c_6 \to +i \gamma c_1}
$
\normalsize

\end{itemize}

Combining $\bs{\Sigma}_1$-$\bs{\Sigma}_5$ gives the following constraints for the mean-field parameters:
\small
\be
\boxed{
\begin{array}{c}
\forall \nu: ~~ q_\nu = 0,~~p_\nu = p, \\
\delta_{12} = -\delta_{23}=\delta_{34}=-\delta_{45}=\delta_{56}=\gamma \delta_{61},\\
m_{12}=m_{23}=m_{34}=m_{45}=m_{56}=-\gamma m_{61}\equiv m
\end{array}
}
\ee
\normalsize

\vspace*{-0.2cm}
\subsection{The mean field parameter $m$}
\vspace*{-0.2cm}
The numerical, self-consistent treatment of the decoupled spin-wave Hamiltonian gives a vanishing mean-field parameter $m$. 
This result does not arise from symmetry and is true only in the asymptotic large-$S$ limit.
For general $S$, $m$ is a very small number.
To see this we consider the self-consistent mean-field Hamiltonian for a single hexagon, 
that corresponds to the decoupled semiclassical problem that we are dealing with:
\small
\be
\mc{H}_{\text{MF}} = 
-h_{\text{loc}} (S_1^w+S_2^w+S_3^w+S_4^w+S_5^w+S_6^w)
+ (S_1^u S_2^u+S_2^v S_3^v+S_3^u S_4^u+S_4^v S_5^v+S_5^u S_6^u-\gamma S_6^v S_1^v) 
\equiv - h_{\text{loc}} S^w_{\text{tot}}+\mc{V}~,
\ee
\normalsize
where $h_{\text{loc}}$ is the self-consistent field exerted from neighboring hexagons and we have taken $K=1$ without loss of generality.  
In what follows we shall use the N\'eel operator  $\mc{L}$ defined as
\small
\be
\mc{L} = S_1^w-S_2^w+S_3^w-S_4^w+S_5^w-S_6^w\,,
\ee
\normalsize
and the relations
\small
\be
[S_1^+ S_2^-, \mc{L}] = [S_1^+ S_2^-, S_1^w-S_2^w] = -2 S_1^+ S_2^- 
\Rightarrow 
\langle g | [S_1^+ S_2^-, \mc{L}]  |g\rangle = -2 \langle g |  S_1^+ S_2^-  |g\rangle \,,
\ee
\normalsize

\begin{itemize}
\item
For $S=1/2$, the numerical, self-consistent solution gives $h_{\text{loc}}=0.37888$ and $m=0$. However, this relation is special to $S=1/2$ because the numerical, self-consistent ground state $|g\rangle$ of $\mc{H}_{\text{MF}}$ has the special property $\mc{L} | g\rangle = 0$. And according to the above relations, this implies that $\langle g |  S_1^+ S_2^-  |g\rangle=0$, which is equivalent with $m=0$.

\item
For $S=1$ and higher, the ground state does not obey the property $\mc{L} | g\rangle = 0$ and $m$ is therefore finite. The numerical solution for $S=1$ gives $h_{\text{loc}}=0.83643$ and $m=0.0011412$, which is a very small number. 

\item In the large-$S$ limit, the parameter $m$ must eventually vanish (consistent with the numerical results from the decoupled, large-$S$ spin-wave Hamiltonian). 
The reason behind this is that as we increase $S$, the ground state $|g\rangle$ comes closer and closer to the classical vacuum $|0\rangle$ (with spins fully polarized along their local ${\bf w}$-axes), which has the property $\mc{L}|0\rangle=0$ (because $|0\rangle$ is an eigenstate of each $S_\nu^w$ individually). 
In fact, this relation remains true when we include the leading effect of semiclassical corrections coming from $\mc{V}$. At this leading level, the ground state wavefunction is given by~\cite{Lindgren1974SM}
\small
\be
|g_1\rangle = |0\rangle + \mc{R} \mc{V} |0\rangle \,.
\ee
\normalsize
where $\mc{R}=\frac{1-|0\rangle \langle 0|}{E_0-\mc{H}_0}$ is the usual resolvent operator. 
To show that $\mc{L}|g_1\rangle=0$ we use the fact that $\mc{L}$ commutes with $\mc{H}_0$ (and therefore with $\mc{R}$ as well) and furthermore $\mc{L}|0\rangle=0$. These properties give:
\small
\be
\mc{L} |g_1\rangle = \mc{L} |0\rangle+  \mc{L} \mc{R} \mc{V} |0\rangle 
=  \mc{R} \mc{L}\mc{V} |0\rangle 
=  \mc{R} [\mc{L},\mc{V}] |0\rangle \,.
\ee
\normalsize
We further have: 
\small
\bea
\mc{V} &=& \frac{1}{4}\left(
S_1^+S_2^+
-S_2^+S_3^+
+S_3^+S_4^+
-S_4^+S_5^+
+S_5^+S_6^+
+\gamma S_6^+S_1^+ 
+ h.c.\right) 
\nonumber\\&&
+ \frac{1}{4}\left(
S_1^+S_2^-
+S_2^+S_3^-
+S_3^+S_4^-
+S_4^+S_5^-
+S_5^+S_6^-
-\gamma S_6^+S_1^- 
+ h.c.\right)
\equiv \mc{V}_1+\mc{V}_2\,.
\eea
\normalsize
Using the standard spin commutation relations we find 
\small
\be
[\mc{L},\mc{V}_1] = 0,~~ 
\text{and}~~ 
[\mc{L},\mc{V}]=[\mc{L},\mc{V}_2]= \frac{1}{2}\left(
S_1^+S_2^-+S_2^+S_3^-+\cdots -\gamma S_6^+S_1^-\right)
- h.c.\,,
\ee 
\normalsize
from which it follows that $[\mc{L},\mc{V}]|0\rangle=0$ and therefore $\mc{L}|g_1\rangle=0$.

At higher orders $n>1$, the ground state $|g_n\rangle$ does not satisfy this property (i.e. $\mc{L}|g_n\rangle \neq 0$), and a finite $m$ is therefore expected (as found explicitly for $S=1$ above, by the exact treatment of the equivalent spin Hamiltonian $\mc{H}_{\text{MF}}$). 
Nevertheless, the important point is that $m$ vanishes asymptotically for large $S$, and it is generally a very small number otherwise ($m=0.0011412$ at $S=1$).  

\end{itemize}

\vspace*{-0.25cm}
\subsection{Two-fold degeneracy structure of the spin-wave spectrum}
\vspace*{-0.25cm}
Fig.~5 of the main text shows that the six spin-wave energies organize into three degenerate pairs. The symmetry origin of this degeneracy can be seen by considering the effect of the operation $\bs{\Sigma}_1$ discussed above. 
We have: 
\small
\be
\vec{B} = {\bf S}^{-1} \cdot {\bf C} = 
{\bf g} \cdot {\bf S}^+ {\bf g} \cdot {\bf C} \Rightarrow 
{\bf g} \cdot \vec{B} =  {\bf S}^+ {\bf g} \cdot {\bf C}
\ee
\normalsize
Let us take the first row of this matrix equation:
\small
\bea
b_1 = {\bf X}_1^\ast \cdot \left(
c_1,c_2,c_3,c_4,c_5,c_6,-c_1^+,-c_2^+,-c_3^+,-c_4^+,-c_5^+,-c_6^+
\right)
\eea
\normalsize
Suppose further that ${\bf X}_1=(a_1,a_2,a_3,a_4,a_5,a_6,a_1',a_2',a_3',a_4',a_5',a_6')$. 
Now, if $b_1$ describes an eigenmode, then $\bs{\Sigma}_1\cdot b_1$ is also an eigenmode with the same energy:
\small
\bea
\bs{\Sigma}_1\cdot b_1 &=& 
(a_1,a_2,a_3,a_4,a_5,a_6,a_1',a_2',a_3',a_4',a_5',a_6')^\ast \cdot  \left(
-i c_4,-i c_5, -i c_6, i  \gamma c_1, i  \gamma c_2, i  \gamma c_3,-ic_4^+,-ic_5^+,-ic_6^+,i \gamma c_1^+,i  \gamma c_2^+,i \gamma c_3^+
\right)\nonumber\\
&=& -i 
(- a_4,- a_5,- a_6,a_1,a_2,a_3, a_4', a_5', a_6',-a_1',-a_2',-a_3')^\ast \cdot \left(
\gamma c_1,\gamma c_2,\gamma c_3,c_4,c_5,c_6,-\gamma c_1^+,-\gamma c_2^+,-\gamma c_3^+,-c_4^+,-c_5^+,-c_6^+
\right)\nonumber\\
&\equiv& -i  {{\bf X}_1'}^\ast \cdot \left(
c_1,c_2,c_3,c_4,c_5,c_6,-c_1^+,-c_2^+,-c_3^+,-c_4^+,-c_5^+,-c_6^+
\right)\nonumber
\eea
\normalsize
This means that the eigenvectors corresponding to the positive (or the negative) eigenvalues of ${\bf g}\cdot{\bf M}$ come in pairs:
\bea
&&
{\bf X}_1 = (a_1,a_2,a_3,a_4,a_5,a_6,a_1',a_2',a_3',a_4',a_5',a_6')\nonumber\\
&&
{\bf X}_1'= (- \gamma a_4,- \gamma a_5,- \gamma a_6,a_1,a_2,a_3, \gamma a_4', \gamma a_5', \gamma a_6',-a_1',-a_2',-a_3')\nonumber
\eea
If these two modes are linearly independent they belong to a 2-dimensional irreducible representation of the symmetry group generated by $\bs{\Sigma}_1$-$\bs{\Sigma}_5$. The numerical results show that this is the case for the whole spectrum of the spin-wave Hamiltonian.

\begin{figure}[!b]
\includegraphics[width=0.2\linewidth]{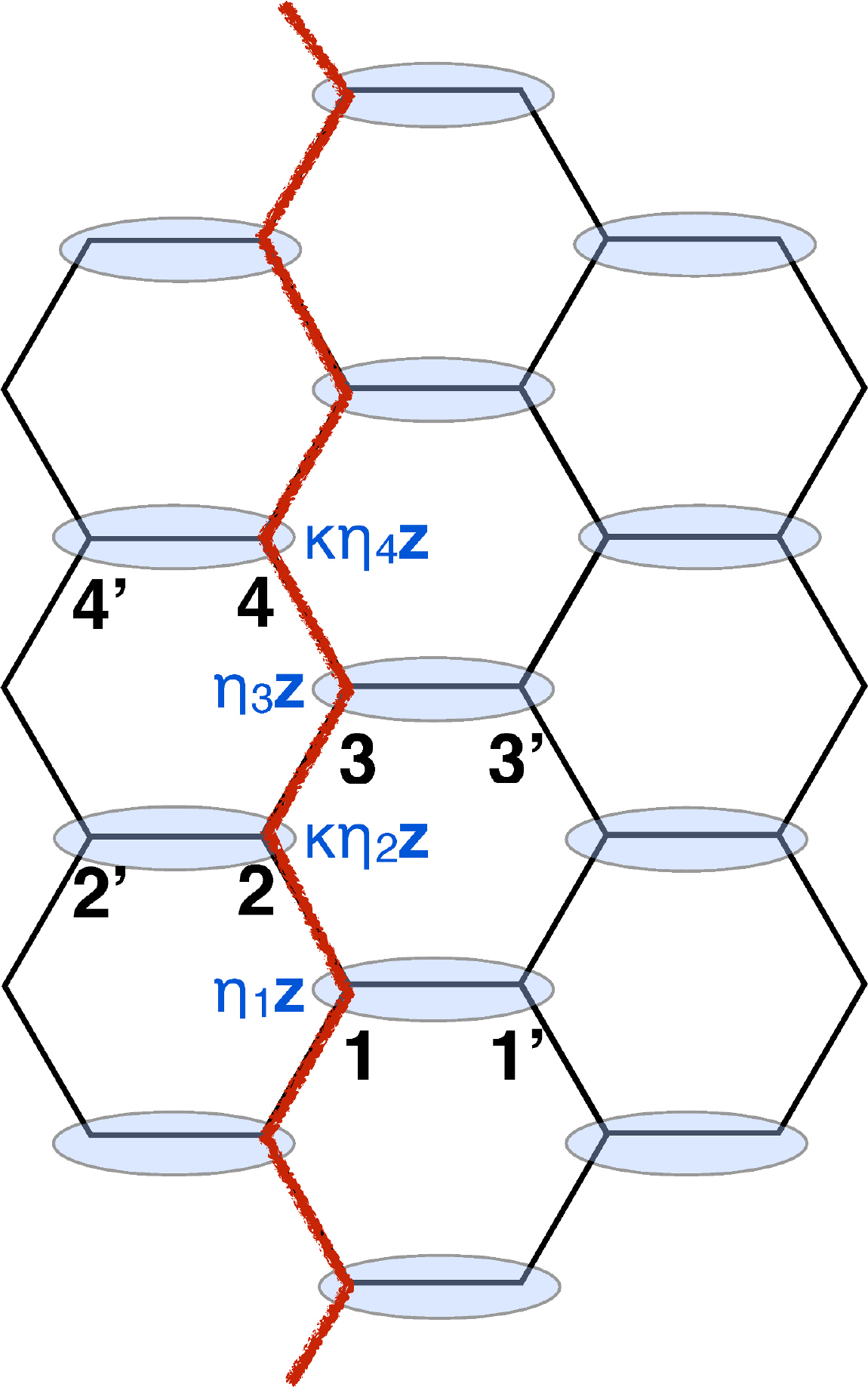}
\caption{\label{fig:ZZstate} Classical states associated with the staggered dimer pattern of Fig.~7 of the main text.} 
\end{figure}

\section{Semiclassical expansion around the states associated with the staggered dimer pattern}
For the numerical data shown in Fig.~8 of the main text we have also performed a non-linear semiclassical expansion around the classical state associated with the staggered pattern of Fig.~7 of the main text. In this pattern, the dimers occupy the horizontal, `zz' bonds, while the empty bonds form infinite strings. Similarly to the above, the strings  decouple from each other and it suffices to consider one string only. To this end we use the labeling scheme of Fig.~\ref{fig:ZZstate} and the following local frames:
\small
\be
\begin{array}{l}
({\bf u}_{1}, {\bf v}_1, {\bf w}_1)=({\bf x}, \eta_1 {\bf y}, \eta_1 {\bf z}),  \\
({\bf u}_2, {\bf v}_2, {\bf w}_2)=({\bf x}, \eta_2 \kappa {\bf y}, \eta_2 \kappa {\bf z}),  \\
({\bf u}_3, {\bf v}_3, {\bf w}_3)=(\eta_2\eta_3 \kappa {\bf x}, \eta_2\kappa  {\bf y}, \eta_3 {\bf z}),  \\
({\bf u}_4, {\bf v}_4, {\bf w}_4)=(\eta_2\eta_3\kappa {\bf x}, \eta_2 \eta_3 \eta_4 {\bf y}, \eta_4\kappa  {\bf z}),  
\end{array}
\ee
\normalsize
and so on. With this choice of local axes we move all the dependence on $\eta$'s on the last bond at infinity. And since the string is infinite, the energy contribution from that last bond does not matter, and therefore the spin wave expansion does not depend on the configuration of $\eta$'s altogether. 
The Hamiltonian for the terms along the string becomes
\small
\bea
\mc{H} = K (S_1^uS_2^u+S_2^vS_3^v+S_3^uS_4^u+ \cdots ) - |K| (S_1^w S_{1'}^w + S_2^w S_{2'}^w+\cdots )  
\eea
\normalsize
The Hamiltonian along the string describe a system with a unit cell of two sites, and we can relabel the sites as follows: 
\small
\be
1\to (R=0,\nu=1),~~ 2\to (R=0,\nu=2),~~ 3\to (R=1,\nu=1), ~~ 4\to (R=1,\nu=2)\,, 
\ee
\normalsize
and so on. 
Keeping only the terms pertaining to the given string and going to momentum space (along the string) gives, in matrix notation:
\small
\be
\mc{H}/|K| = f_0+\frac{1}{2}\sum_k {\bf C}^+_k\cdot {\bf M}_k \cdot {\bf C}_k
\ee
\normalsize
where
\small
\be
{\bf C}_k^+=\left(c_{k,1}^+, c_{k,2}^+,c_{-k,1},c_{-k,2}\right),~~~~
{\bf M}_k =
\left(\begin{array}{c c V{2}  c c}
 d &\chi_{12}(k) & d'& \rho_{12}(k)\\ 
 \chi_{12}(-k)&d & \rho_{12}(-k) &d'\\
 \hlineB{2}
d' &\rho_{12}(k)&d&\chi_{12}(k)\\
\rho_{12}(-k)&d'&\chi_{12}(-k)&d
\end{array}
\right)\,,
\ee
\normalsize
and
\small
\be
\begin{array}{c}
f_0=\big[ p^2 - S^2 - \kappa (\tau_{12} + \tau_{12}') -d\big] N_{\text{s}}/2, ~~~
d=(S-p)-4 \kappa (f_{12}-f_{12}'),~~~d'=-2\kappa(f_{12}+f_{12}'), \\
\rho_{12}(k)=-\kappa (g_{12}+g_{12}' e^{ik}),~~~
\chi_{12}(k)=-\kappa (g_{12}-g_{12}'e^{ik})\,,
\end{array}
\ee
\normalsize
where the constants $f_{12}$, $f'_{12}$, $g_{12}$, $g'_{12}$, $\tau_{12}$ and $\tau'_{12}$ are defined again as in Eqs.~(\ref{eq:fgtau}) and (\ref{eq:fpgptaup}) above, 
and $N_s$ is the number of sites along the string. 
Here, the matrix $\vec{S}_{k}$ must satisfy:
\small
\bea
{\bf S}_k^+ \cdot {\bf g}\cdot {\bf S}_k = {\bf g},~~~
{\bf S}_k = \left(
\begin{array}{cc}
{\bf A}_k & {\bf B}_{-k}^\ast \\
{\bf B}_k & {\bf A}_{-k}^\ast
\end{array}
\right) = {\bf S}_{-k}^\ast\,.
\eea
\normalsize 
Note that the second relation replaces the relation $\bs{L} \cdot {\bf S} \cdot \bs{L} = {\bf S}^\ast$ that we had in Eq.~(\ref{eq:LSL}) above.

\vspace*{-0.25cm}
\subsection{Symmetry constraints}
\vspace*{-0.25cm}
\begin{itemize}
\item 
{\bf Symmetry $\bs{\Sigma}'_1$}. This is a translation by one lattice spacing, followed by a $\pi/2$-rotation around the local ${\bf w}$-axes:
\small
\bea
(S_{R,1}^u,S_{R,1}^v,S_{R,1}^w) \to (S_{R,2}^v,-S_{R,2}^u,S_{R,2}^w),~~~
(S_{R,2}^u,S_{R,2}^v,S_{R,2}^w) \to (S_{R+1,1}^v,-S_{R+1,1}^u,S_{R+1,1}^w)\,, \nonumber
\eea
\normalsize
which is equivalent with \small$\boxed{c_{R,1} \to -i c_{R,2},~~~c_{R,2} \to -i c_{R+1,1}}$.\normalsize

\item 
{\bf Symmetry $\bs{\Sigma}'_2$}. This is a reflection though the bond (2,2') (see Fig.~\ref{fig:ZZstate}), followed by a $\pi/2$-rotation around the local ${\bf w}$-axes:
\small
\bea
(S_{R,1}^u,S_{R,1}^v,S_{R,1}^w) \to (S_{R+1,1}^v,-S_{R+1,1}^u,S_{R+1,1}^w),~~~
(S_{R,2}^u,S_{R,2}^v,S_{R,2}^w) \to (S_{R,2}^v,-S_{R,2}^u,S_{R,2}^w)\,, \nonumber
\eea
\normalsize
which is equivalent with \small$\boxed{c_{R,1} \to -i c_{R+1,1},~~~c_{R,2} \to -i c_{R,2}}$.\normalsize

\item 
{\bf Symmetry $\bs{\Sigma}'_3$}. This is an inversion through the middle of the bond (1,2) (see Fig.~\ref{fig:ZZstate}), which maps $\boxed{c_{R,1} \to c_{R,2}}$. 

\end{itemize}

Combining the symmetries $\bs{\Sigma}'_1$-$\bs{\Sigma}'_3$ gives:
\small
\bea
\boxed{\langle c_{R,1} c_{R,2}\rangle=-\langle c_{R,2} c_{R+1,1}\rangle=\cdots},~~~
\boxed{\langle c_{R,\nu} c_{R,\nu}\rangle=0},~~~
\boxed{\langle c_{R,1} c^+_{R,2}\rangle=\langle c_{R,2} c^+_{R+1,1}\rangle=\cdots}\,.
\eea
\normalsize

%

\end{widetext}

\end{document}